\documentclass[a4paper,11pt]{article}
\pdfoutput=1 

\usepackage{jheppub_modified} 

\usepackage[T1]{fontenc} 

\usepackage{verbatim}
\usepackage{color}
\usepackage{graphicx}
\usepackage{subcaption}

\usepackage{bigcenter}

\usepackage{sectsty}
\allsectionsfont{\boldmath}

\usepackage{xspace}
\usepackage[dvipsnames,table]{xcolor}
\usepackage{tabularx}
\newcolumntype{C}[1]{>{\centering\arraybackslash}p{#1}}

\newcommand{\eq}[1]{Eq.~(\ref{#1})}
\newcommand{\bib}[1]{Ref.~\cite{#1}}
\newcommand{\refs}[1]{Refs.~\cite{#1}}
\newcommand{\bibs}[1]{\cite{#1}}
\newcommand{\fig}[1]{Fig.~\ref{#1}}
\newcommand{\tab}[1]{Table~\ref{#1}}

\newcommand{\sect}[1]{Section~\ref{#1}}
\newcommand{\ssect}[1]{Subsection~\ref{#1}}

\newcommand{\bea}{\begin{eqnarray}}
\newcommand{\eea}{\end{eqnarray}}

\newcommand{\nn}{\nonumber}
\newcommand{\crn}{\nonumber \\}

\newcommand{\fr}{\frac}

\newcommand{\gev}{{\unskip\,\text{GeV}}}

\title{Doubly-polarized $WZ$ hadronic production at NLO QCD+EW: Calculation method and further results}

\author[a]{Duc Ninh Le,}
\author[b]{Julien Baglio,}
\author[a]{Thi Nhung Dao}

\affiliation[a]{Faculty of Fundamental Sciences, PHENIKAA University, Hanoi 12116, Vietnam}
\affiliation[b]{Theoretical Physics Department, CERN, CH-1211 Geneva 23, Switzerland}

\emailAdd{ninh.leduc@phenikaa-uni.edu.vn}
\emailAdd{julien.baglio@cern.ch}
\emailAdd{nhung.daothi@phenikaa-uni.edu.vn}

\preprint{CERN-TH-2022-110}     

\abstract{The doubly-polarized production of $W^\pm Z$ pairs at
  the Large Hadron Collider (LHC) is presented at next-to-leading
  order (NLO) accuracy both for the electroweak (EW) and QCD
  corrections, including a detailed description of the calculational
  method using the double-pole approximation. Numerical results at the
  13 TeV LHC are presented in particular for the $W^- Z$ case in the
  $e^-\bar{\nu}_e \mu^+\mu^-$ channel using ATLAS fiducial cuts and
  for polarized distributions defined in the $WZ$ center-of-mass
  system. The NLO EW corrections relative to the NLO QCD predictions
  are found to be smaller than $5\%$ in most kinematic distributions,
  but can reach the level of $10\%$ in some distributions such as 
  lepton transverse momentum distributions or rapidity separation
  between the electron and the $Z$ boson. EW corrections are not
  uniform for different polarizations. A comparison between the new
  ATLAS measurement of polarization fractions to our theoretical
  prediction is presented.
}

\begin{document}
\maketitle
\flushbottom

\section{Introduction}
\label{sect:intro}

The Standard Model (SM) of particle physics is the current framework
which describes how matter is organized at the most fundamental
level. It allows for a consistent description of the interactions
between quarks and leptons via the exchange of gauge bosons, in
particular the $W$ and $Z$ bosons that are mediators of the
electroweak (EW) interaction. The CERN Large Hadron Collider (LHC) has
produced lots of such EW gauge bosons since the beginning of
its operation, and the amount of collected data in both run I and run
II has allowed for a detailed study of the properties of the $W$ and
the $Z$ bosons. It is expected that with run III starting in 2022,
theorists and experimentalists will be able to search for more
potential new-physics effects in the tail of gauge boson differential
distributions, as well as access to a precise determination of
polarization observables of the $W$ and $Z$ bosons. The four-lepton
channel via $ZZ$ production and the three-lepton channel via $WZ$
productions are of prime importance for such polarization studies, see
for example the latest ATLAS~\cite{ATLAS:2021wob} and CMS
results~\cite{CMS:2021icx} in the three-lepton channel, using the
full run II dataset, as well as ATLAS differential results in the
four-lepton channel~\cite{ATLAS:2021kog}.

In order to allow for a meaningful comparison with experiments, the
theory prediction has to reach a high accuracy and higher order QCD
and EW corrections are needed. The next-to-leading order
(NLO) QCD corrections in the $WZ$ channel were calculated in
\refs{Ohnemus:1991gb,Frixione:1992pj} for on-shell production and in
\refs{Dixon:1998py,Dixon:1999di} for off-shell production including
the leptonic decays. The NLO EW corrections were calculated in
\refs{Accomando:2004de,Bierweiler:2013dja,Baglio:2013toa,
  Biedermann:2017oae}, demonstrating the importance of the
quark-photon real corrections in an inclusive setup. The
next-to-next-to-leading order (NNLO) QCD corrections were obtained for
the first time in
2015~\cite{Gehrmann:2015ora,Grazzini:2016swo,Grazzini:2017ckn} while
the combination of NNLO QCD corrections with NLO EW corrections was
performed in 2019~\cite{Grazzini:2019jkl}. The comparison of theory
predictions with experimental observables also requires soft gluon
effects to be taken into account. This was performed in
\refs{Melia:2011tj,Nason:2013ydw} at NLO QCD in parton shower
programs and then later extended to a consistent matching of NLO
QCD+EW corrections to parton shower in \bib{Chiesa:2020ttl}. In order
to study new physics effects, the effect of SM effective theory
operators was included at NLO QCD+EW via anomalous couplings in
\bib{Chiesa:2018lcs} and taken into account in parton shower programs
at NLO QCD in \refs{Baglio:2019uty,Baglio:2020oqu}. Full NLO QCD
predictions including full off-shell and spin-correlation effects for
leptonic final states can now be easily obtained with the help of
computer tools such as {\tt
  MCFM}~\cite{Campbell:1999ah,Campbell:2011bn} or {\tt
  VBFNLO}~\cite{Arnold:2008rz,Baglio:2014uba}.

All this progress in the calculation of higher-order corrections has
also helped to increase the accuracy of the theoretical predictions
for angular observables and polarized cross sections. Thanks to the
wealth of data at the LHC it is now possible to also get experimental
results for polarization observables in the three- and four-lepton
channels. Even the two-lepton channel has been given some attention,
see the NNLO QCD theory perspective in \bib{Poncelet:2021jmj}. As the
two-lepton channel is more difficult to measure because of the amount
of missing energy, this paper will focus on the three-lepton channel
using $WZ$ production. 

Experimental results for singly-polarized observables in the three-lepton channel were first 
presented by ATLAS in 2019~\cite{ATLAS:2019bsc} using run II data at 13 TeV. They
have been updated very recently including a measurement, for the first
time, of the double polarization of $WZ$ events, in particular
involving longitudinally polarized gauge bosons~\cite{ATLAS:2022conf53}. The LO
prediction for polarized cross sections were presented in the
80s~\cite{Bilchak:1984gv,Willenbrock:1987xz} while the NLO QCD
corrections were studied much later~\cite{Stirling:2012zt}. 
The NLO EW corrections to gauge boson polarization observables in the $WZ$
channel were first calculated in \refs{Baglio:2018rcu,Baglio:2019nmc} 
and combined with QCD corrections. 
The idea of this study is to define polarization observables, which we termed 
{\it fiducial} polarizations, directly from lepton angular distributions. 
The same idea has been implemented in \refs{Rahaman:2018ujg,Rahaman:2019lab}, using polarization asymmetries 
to constrain anomalous triple gauge boson couplings. 
The advantage of this approach is that the observables can be easily calculated at any order 
in perturbation theory with arbitrary kinematic cuts on the leptons using available calculations 
for unpolarized cross sections. 
This has been shown to work for single polarizations. Whether similar observables can be defined for 
double polarization is still an open question. 
The negative side of this method is that the values of those observables depend strongly on the 
lepton cuts, hence can be very different from the values obtained using the on-shell gauge boson approximation.   

The traditional method to define gauge boson polarization is using the
on-shell approximation, allowing for a separation of the 
gauge-boson polarizations at the amplitude level. Following this path,
doubly-polarized predictions for the two-lepton channel
  ($W^+W^-$)~\cite{Denner:2020bcz}, three-lepton channel ($W^\pm
  Z$)~\cite{Denner:2020eck}, and four-lepton channel
  ($ZZ$)~\cite{Denner:2021csi} were obtained at NLO in QCD in 2020 and
  2021 using the double-pole approximation (DPA). The calculation in
  the $ZZ$ channel includes EW corrections as well, see \bib{Denner:2021csi}.
It is worth noticing that, because this polarization separation is 
done at the amplitude level (in contrast to the above fiducial
polarizations which are defined at the cross section level),
it requires a careful definition of the on-shell momenta
and other technical details. The nice thing of this approach is that
it allows for generation of fully polarized events.

Inspired by these results, we have extended their method to cover the
three-lepton channel ($W^\pm Z$), 
where a new ingredient must be added to deal with the photon radiation
off the intermediate on-shell $W$ boson. In \cite{Le:2022lrp} we have
already presented our first results at the NLO QCD+EW
level for the $W^+ Z$ production using the same fiducial 
cuts and reference frame as ATLAS~\cite{ATLAS:2019bsc}. 
The goal of the paper is to present the complete description of the
calculation method behind the results of \bib{Le:2022lrp}, as well as
give results for the $W^- Z$ channel. We will also perform a
comparison between our NLO QCD+EW predictions and the new ATLAS
measurement \cite{ATLAS:2022conf53} for the doubly-polarized cross
sections.

The paper is organized as follows. The definition of polarizations is
given in \sect{sect:pol_def}. The details of the calculation of the
QCD corrections are briefly given in \sect{sect:nlo_qcd} while the EW
corrections are explained in depth in \sect{sect:nlo_ew}, first with
the description of the method to calculate the EW corrections to the
production part in \ssect{sect:nlo_ew:prod}, and then to the decay
part in \ssect{sect:nlo_ew:decay}. The numerical results
  using the ATLAS fiducial cuts and the $WZ$ center-of-mass system (c.m.s)
  reference frame at the LHC at 13 TeV, mainly for the $W^-Z$ channel,
  are presented in \sect{sect:results}. This section starts with the
integrated polarized  cross sections that are given in
\ssect{sect:XS}, where the comparison with the ATLAS measurement is
provided. The relevant kinematical distributions are then shown in
\ssect{sect:dist}. We conclude in \sect{sect:conclusion}.

\section{Definition of polarizations}
\label{sect:pol_def}

\begin{figure}[ht!]
  \centering
  \includegraphics[height=5cm]{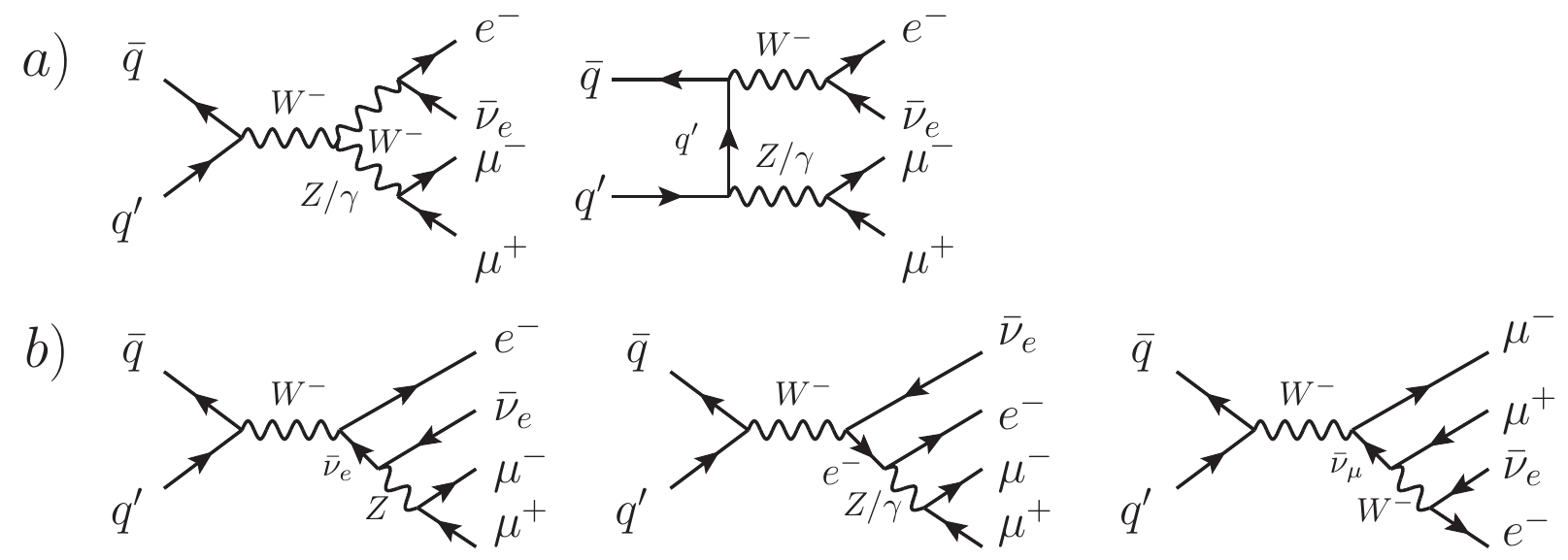}
  \caption{Double and single resonant diagrams at leading order. Group
    a) includes both double and single resonant diagrams, while group
    b) is only single resonant.}
  \label{fig:LO_diags}
\end{figure}

We study the polarized production of three charged leptons plus
missing energy at hadron colliders, so that the process of interest is
\bea
p(k_1) + p(k_2) \to \ell_1(k_3) + \ell_2(k_4) + \ell_3
(k_5) + \ell_4(k_6) + X,
\label{eq:proc1}
\eea
where the final-state leptons can be either $e^+\nu_e\mu^+\mu^-$ or 
$e^-\bar{\nu}_e\mu^+\mu^-$. Representative Feynman diagrams at leading
order (LO) are depicted in \fig{fig:LO_diags}. They can be divided in
two categories: either the upper row of \fig{fig:LO_diags} where the
intermediate $W$ and $Z$  bosons (or photon) both split into
final-state leptons, or the lower row of \fig{fig:LO_diags} which
contains only singly-resonant diagrams with $W\to 4\ell$ splitting. We
are interested in the kinematical region where the final-state leptons
originate from nearly on-shell (OS) gauge bosons, so that the process
in \eq{eq:proc1} can be seen as
\bea
p(k_1) + p(k_2) \to V_1(q_1) + V_2(q_2) \to \ell_1(k_3) + \ell_2(k_4) + \ell_3
(k_5) + \ell_4(k_6) + X,
\label{eq:proc1_VV}
\eea
where the intermediate gauge bosons are $V_1=W^\pm$, $V_2=Z$. In
practice this means we are neglecting the diagrams depicted in the
lower row of \fig{fig:LO_diags} as well as the photon contribution of
the upper row of \fig{fig:LO_diags}. In this way we will be able to
define polarized cross sections and get a clear separation of the
polarizations of the intermediate gauge bosons. This is called the
double-pole approximation (DPA), where contributions from $W\to 4\ell$
and very off-shell $WZ$ contributions are neglected because they are
strongly suppressed by the kinematical cuts. In the DPA, the whole
process can be viewed as an OS production of $WZ$ followed by the OS
decays $W^\pm\to e^\pm \nu_e$ and $Z\to\mu^+\mu^-$, connected via
off-shell $W$ and $Z$ propagators and keeping the spin correlations.

The (unpolarized) amplitude for the process in \eq{eq:proc1} is
defined at LO in the DPA as (see also \bib{Denner:2000bj} for
$e^+e^-\to W^+W^-\to 4$ fermions)
\bea
\mathcal{A}_\text{LO,DPA}^{\bar{q}q'\to V_1V_2\to 4l} = \fr{1}{Q_1Q_2}
\sum_{\lambda_1,\lambda_2=1}^{3}
\mathcal{A}_\text{LO}^{\bar{q}q'\to V_1V_2}(\hat{k}_i)\mathcal{A}_\text{LO}^{V_1\to
    \ell_1\ell_2}(\hat{k}_i)\mathcal{A}_\text{LO}^{V_2\to \ell_3\ell_4}(\hat{k}_i)
,\label{eq:LO_DPA}
\eea
with 
\bea
Q_j = q_j^2 - M_{V_j}^2 + iM_{V_j}\Gamma_{V_j}\, (j=1,2),
\label{eq:Qi_def}
\eea
where $q_1 = k_3+k_4$, $q_2 = k_5 + k_6$, $M_V$ and $\Gamma_V$ are the
physical mass and width of the gauge bosons, and $\lambda_j$ are the
polarization indices of the gauge bosons. A similar factorization
in the terms of the sum holds at higher orders in perturbation
theory. It is crucial that all helicity amplitudes $\mathcal{A}$ in
the numerator are calculated using OS momenta $\hat{k}_i$ for the
final-state leptons as well as OS momenta $\hat{q}_j$ for the
intermediate gauge bosons, derived from the off-shell (full process)
momenta $k_i$ and $q_j$, in order to ensure that gauge invariance in
the amplitudes is preserved. An OS mapping is used to obtain the OS
momenta $\hat{k}_i$ from the off-shell momenta $k_i$. 
This OS mapping is not unique, however it
is known that the induced shift by different mappings is of order
$\alpha \Gamma_V/(\pi M_V)$ \cite{Denner:2000bj}. The next sections
will provide the explicit details on the mappings used in our
calculation.

The sum over the polarization index $\lambda_j$ for a given gauge
boson runs from 1 to 3, because a massive gauge boson has three
physical polarization states: two transverse states $\lambda = 1$ and
$\lambda = 3$ (left and right) and one longitudinal state
$\lambda = 2$. In total the diboson system has then 9 polarization
states, each of them can be singled out of \eq{eq:LO_DPA} by selecting
only the desired $\lambda_j$ in the sum. We will define four main
doubly-polarized cross sections in this paper:
\begin{itemize}
\item The longitudinal-longitudinal (LL, or $W^\pm_L Z^{}_L$)
  contribution, obtained with selecting only $\lambda_1=\lambda_2 = 2$
  in the sum of \eq{eq:LO_DPA};
\item The transverse-transverse (TT, or $W^\pm_T Z^{}_T$)
  contribution, obtained with selecting only $\lambda_1\neq 2$ and
  $\lambda_2\neq 2$ in the sum of \eq{eq:LO_DPA}, taking into account
  the interference between the various individual transverse
  polarization states of the two gauge boson;
\item The longitudinal-transverse (LT, or $W^\pm_L Z^{}_T$)
  contribution, obtained with selecting only $(\lambda_1,\lambda_2) =
  (2,1) + (2,3)$;
\item The transverse-longitudinal (TL, or $W^\pm_T Z^{}_L$)
  contribution, obtained with selecting only $(\lambda_1,\lambda_2) =
  (1,2) + (3,2)$.
\end{itemize}
In addition we will call ``interference'' the difference between the
unpolarized contribution (where all polarization amplitudes are summed
before squaring) and the sum of the LL, TT, LT, and TL cross
sections.

We will present in the next sections our detailed calculation of the
NLO QCD and EW corrections in the DPA and focus on those
doubly-polarized contributions that we have defined above. While the
unpolarized cross section is Lorentz invariant because all possible
helicity states are summed over, the doubly-polarized cross sections
depend on the reference frame. As done in \bib{Le:2022lrp} we will
provide results in the $WZ$ center-of-mass system, following ATLAS
choice in \bib{ATLAS:2019bsc}. The same set of kinematic cuts will be used. 
Note that the NLO QCD corrections have already been presented in
\bib{Denner:2020eck} where the computation method is identical to the one used for
the $WW$ channel~\cite{Denner:2000bj} and for the $ZZ$
production~\cite{Denner:2021csi}. We follow the same steps and
re-describe here the NLO QCD calculation method to prepare the
framework and notations for the NLO EW calculation.

\section{NLO QCD}
\label{sect:nlo_qcd}

The master formulas for
the virtual, real-gluon emission, and quark-gluon induced amplitudes
are schematically written as follows,
\begin{align}
\delta\mathcal{A}_\text{virt,QCD}^{\bar{q}q'\to V_1V_2 \to 4l} &=\fr{1}{Q_1Q_2}
\sum_{\lambda_1,\lambda_2}
\delta\mathcal{A}_\text{virt,QCD}^{\bar{q}q'\to
    V_1V_2}(\hat{k})\mathcal{A}_\text{LO}^{V_1\to
    \ell_1\ell_2}(\hat{k})\mathcal{A}_\text{LO}^{V_2\to
    \ell_3\ell_4}(\hat{k}),\label{eq:virt_QCD}\\
\delta\mathcal{A}_\text{g-rad}^{\bar{q}q'\to V_1V_2g \to 4lg} &=
\fr{1}{Q_1Q_2}\sum_{\lambda_1,\lambda_2}
\delta\mathcal{A}_\text{g-rad}^{\bar{q}q'\to
    V_1V_2g}(\hat{k})\mathcal{A}_\text{LO}^{V_1\to
    \ell_1\ell_2}(\hat{k})\mathcal{A}_\text{LO}^{V_2\to \ell_3\ell_4}(\hat{k}),
\label{eq:rad_QCD}\\
\delta\mathcal{A}_\text{g-ind}^{qg\to V_1V_2q' \to 4l q'} &=
\fr{1}{Q_1Q_2}\sum_{\lambda_1,\lambda_2}
\delta\mathcal{A}_\text{g-ind}^{qg\to
    V_1V_2q'}(\hat{k})\mathcal{A}_\text{LO}^{V_1\to
    \ell_1\ell_2}(\hat{k})\mathcal{A}_\text{LO}^{V_2\to \ell_3\ell_4}(\hat{k}),
\label{eq:ind_QCD}
\end{align}
where the correction amplitudes
$\delta\mathcal{A}_\text{virt,QCD}^{\bar{q}q'\to V_1V_2}$,
$\delta\mathcal{A}_\text{g-rad}^{\bar{q}q'\to V_1V_2g}$, and
$\delta\mathcal{A}_\text{g-ind}^{qg\to V_1V_2q'}$ have been
calculated in the OS production calculation in \bib{Baglio:2013toa}
and are reused here. The amplitudes are for the unpolarized process;
for the polarized amplitudes the corresponding $\lambda_{i,j}$ have to
be selected in the sum.
Note that the amplitude factors on the r.h.s have to be calculated 
using suitable OS mapped momenta denoted by the hat. Details of
the OS mappings are below provided. The propagator factors $Q_i$ are
computed using the off-shell momenta.

For the real corrections \eq{eq:rad_QCD} and \eq{eq:ind_QCD}, since
the amplitudes are divergent in the IR limits we employ the dipole
subtraction method \cite{Catani:1996vz,Dittmaier:1999mb} to calculate
the NLO cross section. In this formalism, the full 
differential cross section is written as, using similar notation as in
\bib{Denner:2021csi}:
\begin{align}
\left(\fr{d\sigma}{d\xi}\right)_\text{NLO} = &\int d\Phi_n^{(4)} \mathcal{B}(\Phi_n^{(4)}) \delta(\xi - \xi_n) \crn
& + \int d\Phi_n^{(4)} \left[ \mathcal{V}(\Phi_n^{(D)}) + \mathcal{C}(\Phi_n^{(D)}) 
+ \int d\Phi_\text{rad}^{(D)} \mathcal{D}_\text{int}(\Phi_n^{(D)},\Phi_\text{rad}^{(D)}) \right]_{D=4} \delta(\xi - \xi_n) \crn
& + \int d\Phi_{n+1}^{(4)} \left[\mathcal{R}(\Phi_{n+1}^{(4)})\delta(\xi - \xi_{n+1}) - \mathcal{D}_\text{sub}(\tilde{\Phi}_n^{(4)},\Phi_\text{rad}^{(4)})\delta(\xi - \tilde{\xi}_{n}) \right],
\label{Xsection_CS_prod}
\end{align}
where $\mathcal{B}$ and $\mathcal{V}$ are the Born and virtual
contributions. The flux factor is included in the matrix elements. 
$\mathcal{R}$ is the real correction term,
which includes the amplitudes $\delta\mathcal{A}_\text{g-rad}$ and
$\delta\mathcal{A}_\text{g-ind}$ above defined. Note that $\xi$ is a placeholder for
any differential variable that is of interest: the $p_T$ of one of the
final-state leptons, the invariant masses, angle, etc. 
$\mathcal{D}_\text{sub}$ is the dipole subtraction term introduced in
the Catani-Seymour (CS) formalism \cite{Catani:1996vz}. For the NLO
QCD, $\mathcal{D}_\text{sub}$ includes only the case of initial-state
emitter and initial-state spectator. The corresponding integrated
$\mathcal{D}_\text{int}$ term is placed in the same group with the
virtual contribution and also depends on the radiation
  phase-space $\Phi_\text{rad}^{(D)}$ (here in $D$ dimensions).
Finally, to cancel the left-over initial state collinear divergences,
the collinear counter term $\mathcal{C}$ has to be added. The tilde
placed on top of $\Phi_n$ and $\xi_n$ is to indicate that the momenta
are calculated using the CS mappings. To regulate the IR divergences
we use as default the mass regularization, i.e. introducing small mass
parameters for the gluon and the light quarks (all but the top
quark). We have verified that the result is in good agreement with the
one obtained using  dimensional regularization. 

For the $\mathcal{R}$ contribution, we first generate a set of $(n+1)$ off-shell 
momenta $[k_{n+1}]$ in the partonic c.m.s. We then perform the
following OS mapping:
\begin{itemize}
\item Boost all momenta to the $VV$ c.m.s\footnote{Note that the partonic
  c.m.s does not always coincide with the VV c.m.s, because of the
  extra real radiation at NLO.};
\item Perform the OS projection on the four lepton momenta. We call 
this OSVV4 mapping (which is the $\text{DPA}^{(2,2)}$ mapping in \cite{Denner:2021csi}),
\bea
[\hat{k}_{n+1}] = \text{OSVV4 mapping}([k_{n+1}]),
\label{eq:OSVV4}
\eea
where VV indicates that this mapping has to be done in the $VV$ c.m.s
and 4 to denote the $VV \to 4\,l$ transition (in the next section we will have $VV
\to 4\,l+\gamma$ for NLO EW final-state radiation). We note that the
$VV$ c.m.s of the new OS momenta $[\hat{k}_{n+1}]$ coincides with the
$VV$ c.m.s of the off-shell momenta, because momentum conservation requires that
\bea
\hat{q}_1 + \hat{q}_2 = q_1 + q_2,\label{eq:mom_conserving_qi}
\eea 
where $q_{1,2}$ are the momenta of the gauge bosons.
\end{itemize}

To be more explicit, the OSVV4 mapping is done as follows \cite{Denner:2021csi}. 
After boosting all momenta to the VV c.m.s, we first calculate the OS momenta $\hat{q}_i$ 
which satisfy \eq{eq:mom_conserving_qi} together with the on-shellness $\hat{q}^2_i = M^2_{V_i}$. 
These conditions are however not enough to fix all the components of $\hat{q}_i$. 
For this, we choose that the spatial direction of the gauge bosons in the VV c.m.s is preserved 
as in \cite{Denner:2000bj}, namely $\vec{\hat{q}}_1 = b\vec{q}_1$ with $b$ being a real number. 
This coefficient $b$ is then easily calculated and the result is provided in \cite{Baglio:2018rcu} 
(see Appendix A). The OS final-state lepton momenta are computed as \cite{Denner:2021csi}: 
\begin{itemize}
\item Boost $k_{e}$ and $k_{\nu_e}$ into the off-shell $W$ boson rest
  frame, calculate the spatial direction $\vec{n}_e$ in this frame;
\item Set the spatial direction of $\hat{k}_{e}^{\prime}$ in the
  on-shell $W$ boson rest frame to be the same as in the off-shell
  $W$ boson rest frame, so that, in the on-shell $W$ boson rest
  frame, we have $\vec{\hat{k}}_e^{\prime}=\vec{n}_e \hat{k}_e^{\prime 0}$ 
  with $\hat{k}_e^{\prime 0} = M_W/2$. For the neutrino $\vec{\hat{k}}_{\nu_e}^{\prime} = -\vec{\hat{k}}_e^{\prime}$ 
  and $\hat{k}_{\nu_e}^{\prime 0} = \hat{k}_e^{\prime 0}$;
\item Boost back the momenta $\hat{k}_l^{\prime}$ from the on-shell
  $W$ rest frame to the VV c.m.s using the
  boot parameters $\hat{q}_1$ to obtain the OS momenta $\hat{k}_{e}$
  and $\hat{k}_{\nu_e}$.
\end{itemize}
The same procedure is applied to the $Z$ decay products, replacing
above the $W$ boson by the $Z$ boson ($q_1$ by $q_2$, $k_e$ and
$k_{\nu_e}$ by $k_{\mu^+}$ and $k_{\mu^-}$). The initial quark momenta are 
unchanged in this mapping.

For the subtraction term $\mathcal{D}_\text{sub}$, we first calculate
the CS momenta (also called the CS reduced momenta) $[\tilde{k}_{n}]$
from the off-shell momenta $[k_{n+1}]$ (generated in the partonic
c.m.s) using the CS mapping for the case of initial-state emitter and
initial-state spectator as in \bib{Dittmaier:1999mb} (massless case):
\bea
[\tilde{k}_{n}] = \text{CS mapping}([k_{n+1}]).
\eea
We then boost $[\tilde{k}_{n}]$ to the $VV$ c.m.s and perform the
OSVV4 mapping to obtain a set of OS momenta:
\bea
[\hat{\tilde{k}}_{n}] = \text{OSVV4 mapping}([\tilde{k}_{n}]).
\eea
The DPA amplitudes for the subtraction terms are calculated similarly
to \eq{eq:LO_DPA} with the OS momenta $[\hat{\tilde{k}}_{n}]$ entering
the amplitudes in the numerator and the off-shell momenta
$[\tilde{k}_{n}]$ in the denominator factors $Q_j$. We note that $Q_j$
are Lorentz invariant hence it does not matter in which frame they are
calculated. However, the helicity amplitudes in the numerator are not
Lorentz invariant, hence they have to be calculated in the correct
frame, which is the VV c.m.s in our case.

For kinematic cuts and distributions, we use the off-shell momenta
$[k_{n+1}]$ for the $\mathcal{R}$ term and the CS mapped momenta
$[\tilde{k}_{n}]$ for the $\mathcal{D}_\text{sub}$ term. These momenta
have to be boosted from the partonic c.m.s to the Lab frame before
applying cut constraints or filling histograms.

Concerning the integrated dipole terms, they are calculated as in
\cite{Dittmaier:1999mb} with the only addition that the OS mapping
step has to be implemented when calculating the Born amplitudes. Since
the same on-shell mapping is used in the $\mathcal{D}_\text{sub}$ term
and in its integrated counterpart $\mathcal{D}_\text{int}$, the needed
correspondence between these two terms are guaranteed.

\section{NLO EW}
\label{sect:nlo_ew}

NLO EW corrections in the DPA are divided into production and decay
parts. For the production part, full NLO EW corrections to the process
$\bar{q}q'\to WZ$ are calculated. For the decay part, full NLO EW
corrections to the decays $W\to e \nu_e$ and $Z\to \mu^+ \mu^-$  are
included. These are called factorizable corrections. The
non-factorizable contribution, including interferences between the
initial-state radiation and the final-state radiation, as defined in
\cite{Baglio:2018rcu}, is very
small~\cite{Beenakker:1997bp,Denner:1997ia,Denner:1998rh} and hence
neglected.

For the production part, the NLO amplitudes read
\begin{align}
\delta\mathcal{A}_\text{virt,prod}^{\bar{q}q'\to V_1V_2 \to 4l} &=\fr{1}{Q_1Q_2}
\sum_{\lambda_1,\lambda_2}
\delta\mathcal{A}_\text{virt,prod}^{\bar{q}q'\to
    V_1V_2}\mathcal{A}_\text{LO}^{V_1\to
    \ell_1\ell_2}\mathcal{A}_\text{LO}^{V_2\to
    \ell_3\ell_4},\label{eq:virt_EW_prod}\\
\delta\mathcal{A}_\text{$\gamma$-rad,prod}^{\bar{q}q'\to V_1V_2 \to 4l\gamma} &=\fr{1}{Q_1Q_2}
\sum_{\lambda_1,\lambda_2}
\delta\mathcal{A}_\text{$\gamma$-rad,prod}^{\bar{q}q'\to
    V_1V_2\gamma}\mathcal{A}_\text{LO}^{V_1\to
    \ell_1\ell_2}\mathcal{A}_\text{LO}^{V_2\to \ell_3\ell_4},
\label{eq:rad_EW_prod}\\
\delta\mathcal{A}_\text{$\gamma$-ind,prod}^{q\gamma\to V_1V_2q' \to 4l q'} &=
\fr{1}{Q_1Q_2}\sum_{\lambda_1,\lambda_2}
\delta\mathcal{A}_\text{$\gamma$-ind,prod}^{q\gamma\to
    V_1V_2q'}\mathcal{A}_\text{LO}^{V_1\to
    \ell_1\ell_2}\mathcal{A}_\text{LO}^{V_2\to \ell_3\ell_4},
\label{eq:ind_EW_prod}
\end{align}
where the correction amplitudes
$\delta\mathcal{A}_\text{virt,prod}^{\bar{q}q'\to V_1V_2}$,
$\delta\mathcal{A}_\text{$\gamma$-rad,prod}^{\bar{q}q'\to V_1V_2\gamma}$, and
$\delta\mathcal{A}_\text{$\gamma$-ind,prod}^{q\gamma\to V_1V_2q'}$ have been
calculated in the OS production calculation in \bib{Baglio:2013toa}
and are reused here. 

For the $W$ decay part, we have
\begin{align}
\delta\mathcal{A}_\text{virt,W}^{\bar{q}q'\to V_1V_2 \to 4l} &=\fr{1}{Q_1Q_2}
\sum_{\lambda_1,\lambda_2}
\mathcal{A}_\text{LO}^{\bar{q}q'\to
    V_1V_2}\delta\mathcal{A}_\text{virt}^{V_1\to
    \ell_1\ell_2}\mathcal{A}_\text{LO}^{V_2\to
    \ell_3\ell_4},\label{eq:virt_EW_decayW}\\
\delta\mathcal{A}_\text{$\gamma$-rad,W}^{\bar{q}q'\to V_1V_2 \to 4l\gamma} &=\fr{1}{Q'_1Q_2}
\sum_{\lambda_1,\lambda_2}
\mathcal{A}_\text{LO}^{\bar{q}q'\to
    V_1V_2}\delta\mathcal{A}_\text{$\gamma$-rad}^{V_1\to
    \ell_1\ell_2\gamma}\mathcal{A}_\text{LO}^{V_2\to \ell_3\ell_4},
\label{eq:rad_EW_decayW}
\end{align}
and for the $Z$ decay: 
\begin{align}
\delta\mathcal{A}_\text{virt,Z}^{\bar{q}q'\to V_1V_2 \to 4l} &=\fr{1}{Q_1Q_2}
\sum_{\lambda_1,\lambda_2}
\mathcal{A}_\text{LO}^{\bar{q}q'\to
    V_1V_2}\mathcal{A}_\text{LO}^{V_1\to
    \ell_1\ell_2}\delta\mathcal{A}_\text{virt}^{V_2\to
    \ell_3\ell_4},\label{eq:virt_EW_decayZ}\\
\delta\mathcal{A}_\text{$\gamma$-rad,Z}^{\bar{q}q'\to V_1V_2 \to 4l\gamma} &=\fr{1}{Q_1Q'_2}
\sum_{\lambda_1,\lambda_2}
\mathcal{A}_\text{LO}^{\bar{q}q'\to
    V_1V_2}\mathcal{A}_\text{LO}^{V_1\to
    \ell_1\ell_2}\delta\mathcal{A}_\text{$\gamma$-rad}^{V_2\to \ell_3\ell_4\gamma},
\label{eq:rad_EW_decayZ}
\end{align}
where the NLO decay amplitudes are generated by {\tt FormCalc}
\cite{Hahn:1998yk,Hahn:2000kx}. The new variables $Q'_1$ and $Q'_2$
are defined as in \eq{eq:Qi_def} but with the gauge-boson momenta
being reconstructed from the off-shell $1 \to 3$ decays. 

As in the case of NLO QCD, the amplitude factors on the r.h.s of 
Eqs.~(\ref{eq:virt_EW_prod}-\ref{eq:rad_EW_decayZ}) 
are calculated using on-shell momenta, while the factors $Q_i$ and
$Q'_i$ are off-shell.

\subsection{NLO EW corrections to the production part}
\label{sect:nlo_ew:prod}

The differential cross section for the production part is calculated
using the same formalism as in \eq{Xsection_CS_prod}. The main
differences compared to the NLO QCD case come from the real
corrections because the photon can now be radiated off the on-shell
$W$ boson, leading to new types of dipole terms in the CS subtraction
function.

We split the $\mathcal{R}$ term in \eq{Xsection_CS_prod} into two parts
\bea
\mathcal{R}^\text{prod} = \mathcal{R}_\text{$\gamma$-ind}^\text{prod} + \mathcal{R}_\text{$\gamma$-rad}^\text{prod},
\eea
corresponding to two subtraction terms
$\mathcal{D}_\text{$\gamma$-ind}^\text{prod,sub}$ and
$\mathcal{D}_\text{$\gamma$-rad}^\text{prod,sub}$, respectively. 

For the $\mathcal{R}_\text{$\gamma$-ind}^\text{prod}$ and
$\mathcal{R}_\text{$\gamma$-rad}^\text{prod}$ terms, the OS momenta
$[\hat{k}_{n+1}]$ are calculated using the same method as in the NLO
QCD case. Concerning the dipole subtraction term ($\mathcal{D}_\text{sub}$)
of the $\gamma$-induced process, the same method as in the NLO QCD
case is used, meaning that the OS mapping is applied on top of the CS
reduced momenta. The integrated counterpart is therefore treated
accordingly with the same OS mapping. We note that the $\gamma \to
W^+W^-$ splitting is finite due to the $W$ mass hence there is no
subtraction term for this splitting. The Born amplitude in the
subtraction term is therefore proportional to the $\bar{q}q'\to V_1V_2
\to 4l$ amplitude as in the NLO QCD case.

For the dipole subtraction term ($\mathcal{D}_\text{sub}$)
of the $\gamma$-radiated process, there are two contributions: 
both emitter and spectator are initial-state particles or 
one is in the initial state the other is the OS $W$ boson. 
The latter one is needed, even though the $W$ is an intermediate particle, 
because the OS amplitude $\delta\mathcal{A}_\text{$\gamma$-rad,prod}^{\bar{q}q'\to V_1V_2\gamma}$ 
in \eq{eq:rad_EW_prod} contains soft divergences due to the photon radiation off an OS $W$. 
The case of initial-state emitter and initial-state spectator is treated 
identically to the NLO QCD process. 

For the case of $W$ emitter and initial quark spectator, which is the 
case of final-state emitter and initial-state spectator in \cite{Dittmaier:1999mb}, 
the subtraction function for the OS production $\bar{q} q' \to W Z
\gamma$ reads \cite{Dittmaier:1999mb}
\begin{align}
\mathcal{\hat{D}}_\text{sub}^{Wq}(\hat{k}) &\sim \hat{g}_\text{sub}(\hat{k}_a,\hat{k}_W,\hat{k}_\gamma)
\mathcal{\hat{B}}(\tilde{\hat{k}}_q,\tilde{\hat{k}}_V),\\
\hat{g}_\text{sub} &= \fr{1}{(\hat{k}_W \hat{k}_\gamma)\hat{x}_{ia}}\left(\fr{2}{2-\hat{x}_{ia}-\hat{z}_{ia}} - 1 - \hat{z}_{ia} - \fr{M_W^2}{\hat{k}_W \hat{k}_\gamma} \right),\\
\hat{x}_{ia} &= \fr{\hat{k}_a \hat{k}_W + \hat{k}_a \hat{k}_\gamma - \hat{k}_W \hat{k}_\gamma}{\hat{k}_a \hat{k}_W + \hat{k}_a \hat{k}_\gamma}, \quad \hat{z}_{ia} = \fr{\hat{k}_a \hat{k}_W}{\hat{k}_a \hat{k}_W + \hat{k}_a \hat{k}_\gamma},\\
\tilde{\hat{k}}_W &= \hat{k}_W + \hat{k}_\gamma - (1-\hat{x}_{ia})\hat{k}_a, \;\; \tilde{\hat{k}}_a = \hat{x}_{ia} \hat{k}_a,
\end{align}
where the subscript $a$ denotes the initial spectator, $i$ denotes the
final emitter and the remaining momenta $[\tilde{\hat{k}}]$ are the
same as the corresponding momenta $[\hat{k}]$. The OS Born amplitude
is $\mathcal{\hat{B}}$. When including the leptonic decays in the 
DPA framework we have
\bea
\mathcal{D}_\text{sub}^{Wq}(k)\delta(\xi-\tilde{\xi}_n) \sim \hat{g}_\text{sub}(\hat{k}_a,\hat{k}_W,\hat{k}_\gamma)
\mathcal{B}(\hat{\tilde{k}}_n,\tilde{Q}_i)\delta(\xi-\tilde{\xi}_n),
\label{dipole_sub_Wq}
\eea
where the OS momenta $[\hat{k}_{n+1}]$ are calculated from the
off-shell $[k_{n+1}]$ as follows.
\begin{itemize}
\item Boost all lepton momenta from the partonic c.m.s to the $VV$ c.m.s;
\item Perform the OS projection on the four lepton momenta as in \eq{eq:OSVV4};
\item Boost all lepton momenta back to the partonic c.m.s.
\end{itemize}
We note that the singular function $\hat{g}_\text{sub}$ is Lorentz
invariant and hence can be calculated in any reference frame. However,
the helicity amplitude factor
$\mathcal{B}(\hat{\tilde{k}}_n,\tilde{Q}_i)$ is not Lorentz invariant
when calculating individual gauge-boson polarizations. It is therefore
important in which frame the momenta $[\hat{\tilde{k}}_n]$ are
calculated. In order to calculate $[\hat{\tilde{k}}_n]$, we first need to 
compute $[\tilde{k}_n]$. This is explained next.

In the above OS mapping, the initial state momenta and the photon
momentum are untouched, hence we have $\hat{k}_a = k_a$,
$\hat{k}_\gamma = k_\gamma$. For the reduced amplitude $\mathcal{B}$
we need the off-shell momenta $[\tilde{k}_n]$ for the denominators
$\tilde{Q}_i$ as well as the corresponding on-shell momenta
$[\hat{\tilde{k}}_n]$ for the amplitudes in the numerator. Moreover,
the off-shell lepton momenta are needed for the kinematic cuts. These
momenta are calculated as follows:
\begin{align}
k_W &= k_e + k_{\nu_e},\\
x_{ia} &= \fr{k_a k_W + k_a k_\gamma - k_W k_\gamma}{k_a k_W + k_a k_\gamma}, \quad z_{ia} = \fr{k_a k_W}{k_a k_W + k_a k_\gamma},\\
\tilde{k}_W &= k_W + k_\gamma - (1-x_{ia})k_a, \;\; \tilde{k}_a = x_{ia} k_a,
\end{align}
and for the remaining momenta $\tilde{k}_j = k_j$ with $j=q'$ (the
other initial quark), $\mu^+,\mu^-$ (the $Z$ decay products). We note
that the only change compared to the OS production case is the
replacement from the OS $\hat{k}_W$ to the off-shell $k_W$. It is easy
to check that the condition $\tilde{k}_W^2 = k_W^2$ is maintained by
this mapping. We then need the off-shell momenta for the $W$ decay
products, corresponding to the reduced momentum $\tilde{k}_W$. These
lepton momenta must satisfy:
\bea
\tilde{k}_e + \tilde{k}_{\nu_e} = \tilde{k}_W,\quad \tilde{k}_e^2 = \tilde{k}_{\nu_e}^2 = 0.
\eea
They are calculated as follows.
\begin{itemize}
\item Boost the momenta $k_e$ and $k_{\nu_e}$ from the partonic
  center-of-mass system to the rest frame of $k_W$, calculate the
  spatial directions $\vec{n}_e$ and $\vec{n}_{\nu_e}$ in this frame.
\item The new off-shell lepton energies in the $\tilde{k}_W$ rest
  frame are $\tilde{k}_e^{\prime 0} = \tilde{k}_{\nu_e}^{\prime 0} =
  \sqrt{\tilde{k}_W^2}/2$ (which is equal to $\sqrt{k_W^2}/2$). The
  spatial directions of the new off-shell lepton momenta in the
  $\tilde{k}_W$ rest frame are taken to be the same as the
  corresponding ones in the $k_W$ rest frame,
  i.e. $\vec{n}_e^\prime=\vec{n}_e$ and
  $\vec{n}_{\nu_e}^\prime=\vec{n}_{\nu_e}$. Using the on-shell
  condition for the leptons, all spatial components are then
  calculated, i.e. $\tilde{k}_l^{\prime i}=n_l^i
  \tilde{k}_l^{\prime 0}$ with $i=1,2,3$ and $l=e,\nu_e$.
\item Boost the new momenta $\tilde{k}_l^{\prime}$ from the
  $\tilde{k}_W$ rest frame to the partonic center-of-mass system using
  the boost parameters $\tilde{k}_W$ to obtain the off-shell CS mapped
  momenta $\tilde{k}_l$ for the $W$ decay products.
\end{itemize}
We note that this off-shell mapping is exactly in the same spirit as 
the on-shell mapping OSVV4. With this step, we have successfully
mapped the off-shell momenta $[k_{n+1}]$ to the off-shell momenta
$[\tilde{k}_n]$ satisfying the energy-momentum conservation and
\bea
k_W^2 = (k_e + k_{\nu_e})^2 = (\tilde{k}_e + \tilde{k}_{\nu_e})^2, \quad \tilde{k}_I^2 = 0 \;\; (I=1,n).
\eea
Finally, as in the case of NLO QCD, we boost $[\tilde{k}_n]$ to the
$VV$ c.m.s then apply the OSVV4 mapping to obtain $[\hat{\tilde{k}}_n]$. The
case of initial quark emitter and $W$ spectator is calculated using the same method.

Concerning the integrated dipole terms, since the singular function
$\hat{g}_\text{sub}$ in \eq{dipole_sub_Wq} is calculated using the OS
momenta, the corresponding integrated functions
$\hat{\mathcal{G}}_\text{sub}$ and $\hat{G}_\text{sub}$ provided in
Appendix A.1 of \cite{Dittmaier:1999mb} have to be calculated with the
OS momenta as well (i.e. in Eqs.~(A.2), (A.3), and (A.4) of \bib{Dittmaier:1999mb} we set $m_i^2=M_W^2$ being the 
$W$ OS mass squared and $P_{ia}^2 = (\hat{k}_W - \hat{k}_q)^2$ is the on-shell quantity) 
to maintain the needed correspondence between the
subtraction term and its integrated counterpart. The Born amplitudes
are of course calculated using OS momenta. The other things are
unchanged in comparison to \cite{Dittmaier:1999mb}.

\subsection{NLO EW corrections to the decay part}
\label{sect:nlo_ew:decay}
The $Z$ decay case is calculated following the method described in
\cite{Denner:2021csi}. We sketch here only the important
points. First, the OS mapping $\text{DPA}^{(3,2)}$ defined in
\cite{Denner:2021csi} is used for the $Z \to \mu^+ \mu^- \gamma$ decay
to generate the OS momenta $[\hat{k}_{n+1}]$.
For completeness we recall here the steps of this
mapping $\text{DPA}^{(3,2)}$, reminding that the momenta are
originally defined in the VV c.m.s,
$k_Z=k_{\mu^+}+k_{\mu^-}+k_{\gamma}$, $\hat{k}_Z$ is the $Z$ OS momentum 
calculated as in the NLO QCD part:
\begin{itemize}
\item Boost $k_{\mu^+}$, $k_{\mu^-}$, and $k_{\gamma}$ into the
  off-shell $Z$ boson rest frame, to calculate the spatial directions
  $\vec{n}_{\mu^+}$, $\vec{n}_{\mu^-}$, and $\vec{n}_{\gamma}$;
\item Rescale the lepton and photon energies according to the
  on-shell-ness of the $Z$ boson, so that we rescale $k_{l}^0$ (taken
  in the off-shell $Z$ boson rest frame) by 
  $M_Z/\sqrt{k_Z^2}$ with $l=\mu^+,\mu^-,\gamma$;
\item Set the spatial directions of $\hat{k}_{l}^{\prime}$ in the
  on-shell  $Z$ boson rest frame to be the same as in the off-shell
  $Z$ boson rest frame, so that, in the on-shell $Z$ boson rest
  frame, we have $\hat{k}_l^{\prime i}=n_l^i \hat{k}_l^{\prime
    0}$ with $i=1,2,3$, $l=\mu^+,\mu^-,\gamma$, and $\hat{k}_l^{\prime
    0} = k_{l}^0 M_Z/\sqrt{k_Z^2}$;
\item Boost back the momenta $\hat{k}_l^{\prime}$ from the on-shell
  $Z$ rest frame to the VV c.m.s using the
  boost parameters $\hat{k}_Z$ to obtain the OS momenta $\hat{k}_{l}$,
  $l=\mu^+,\mu^-,\gamma$.
\end{itemize}

We then apply the CS
mapping (the case of final-state emitter and final-state spectator as
defined in \cite{Dittmaier:1999mb}) on $[\hat{k}_{n+1}]$ to obtain
$[\tilde{\hat{k}}_{n}]$. These momenta are on-shell by
definition. Applying the same CS mapping to the corresponding
off-shell momenta $[k_{n+1}]$ gives the off-shell CS mapped momenta
$[\tilde{k}_{n}]$ needed for the $Q_i$ factors and kinematic cuts in
the subtraction term. The same CS mapping can be used for both OS and
off-shell momentum sets because the OS mapping $\text{DPA}^{(3,2)}$
has been designed for this purpose \cite{Denner:2021csi}. For the
singular function $g_{ij}^\text{sub}$ (defined in Eq.~(3.1) of \bib{Dittmaier:1999mb}, emitter $i$, spectator $j$) in
the subtraction term, we use the on-shell momentum,
i.e. $k_Z^2 = M_Z^2$ to match the corresponding on-shell requirement
in the $Z \to \mu^+ \mu^- \gamma$ decay. To maintain the needed
correspondence between the subtraction term and its integrated
counterpart, we must impose $k_Z^2 = M_Z^2$ as well when calculating
the endpoint function $G_{ij}^\text{sub}$ (defined in Eq.~(3.7) of \bib{Dittmaier:1999mb}).

The new thing in this calculation is the $W$ decay where the photon
can be radiated off the $W$ or the electron. It turns out that the
above method for the $Z$ decay works for this case as well. We
implement exactly the same steps. For the subtraction term and its integrated 
counterpart, the CS mapping and the singular functions $g_{ia}^\text{sub}$, $G_{ia}^\text{sub}$ 
are provided in \cite{Basso:2015gca}. 
\section{Numerical results}
\label{sect:results}
We use the same set of input parameters as in \bib{Le:2022lrp}. 
Results will be presented for the LHC at $13$ TeV center-of-mass energy. 
The factorization and renormalization scales are chosen at a fixed value 
$\mu_F = \mu_R = (M_W + M_Z)/2$, where $M_W=80.385\,\gev$ and $M_Z = 91.1876\,\gev$. 
The parton distribution functions (PDF) are computed using the Hessian set 
{\tt
  LUXqed17\char`_plus\char`_PDF4LHC15\char`_nnlo\char`_30}~\bibs{Watt:2012tq,Gao:2013bia,Harland-Lang:2014zoa,Ball:2014uwa,Butterworth:2015oua,Dulat:2015mca,deFlorian:2015ujt,Carrazza:2015aoa,Manohar:2016nzj,Manohar:2017eqh} via the library {\tt LHAPDF6}~\bibs{Buckley:2014ana}.

The electromagnetic coupling is
obtained from the Fermi constant as $\alpha=\sqrt{2}G_F
M_W^2(1-M_W^2/M_Z^2)/\pi$ with $G_{F} = 1.16637\times 10^{-5} \gev^{-2}$. 
Since the widths are needed for the off-shell propagators in \eq{eq:Qi_def}, we use
$\Gamma_W = 2.085\,\gev$ and $\Gamma_Z = 2.4952\,\gev$. 
For the EW corrections we further need $M_t = 173\,\gev$, $M_H=125\,\gev$. 
The leptons and the light quarks, {\it i.e.} all but the top
quark, are approximated as massless. This is justified as the final results are 
insensitive to these small masses. 

For NLO QCD results, an extra parton radiation occurs. This emission is treated 
inclusively and no jet cuts are used. For NLO EW predictions, an additional photon 
can be emitted. Before applying real analysis cuts on the charged leptons, we 
do lepton-photon recombination to define a dressed lepton. A dressed lepton 
is defined as $p'_\ell = p_\ell + p_\gamma$ if $\Delta
R(\ell,\gamma) \equiv \sqrt{(\Delta\eta)^2+(\Delta\phi)^2}< 0.1$, i.e. when the photon 
is close enough to the bare lepton. Here the letter $\ell$ can be either $e$ or $\mu$ 
and $p$ denotes momentum in the Lab frame. 
Finally, the ATLAS fiducial phase-space cuts used in \refs{Aaboud:2016yus,ATLAS:2019bsc,ATLAS:2022conf53} 
are applied on the dressed leptons as follows 
\bea
        p_{T,e} > 20\gev, \quad p_{T,\mu^\pm} > 15\gev, \quad |\eta_\ell|<2.5,\crn
        \Delta R\left(\mu^+,\mu^-\right) > 0.2, \quad \Delta R\left(e,\mu^\pm\right) > 0.3, \label{eq:cut_default}\\
        m_{T,W} > 30\gev, \quad \left|m_{\mu^+\mu^-} - M_Z\right| < 10\gev\, ,\nn
\eea 
where $m_{T,W} = \sqrt{2p_{T,\nu} p_{T,e} [1-\cos\Delta\phi(e,\nu)]}$ with $\Delta\phi(e,\nu)$ being 
the angle between the electron and the neutrino in the transverse plane. 

The results presented in the next sections are mostly for the $W^- Z$ process as the results for the 
$W^+ Z$ channel have been presented in \cite{Le:2022lrp}, except in \tab{tab:compare_ATLAS} and \fig{fig:dist_Dy_eZ} 
where $W^+ Z$ results are shown. 

\subsection{Integrated polarized cross sections}
\label{sect:XS}
\begin{table}[th!]
 \renewcommand{\arraystretch}{1.3}
\begin{bigcenter}
\setlength\tabcolsep{0.03cm}
\fontsize{7.0}{7.0}
\begin{tabular}{|c|c|c|c|c|c|c|c|c|c|}\hline
  & $\sigma_\text{LO}\,\text{[fb]}$ & $f_\text{LO}\,\text{[\%]}$  & $\sigma^\text{EW}_\text{NLO}\,\text{[fb]}$ & $f^\text{EW}_\text{NLO}\,\text{[\%]}$ & $\sigma^\text{QCD}_\text{NLO}\,\text{[fb]}$ & $f^\text{QCD}_\text{NLO}\,\text{[\%]}$ & $\sigma^\text{QCDEW}_\text{NLO}\,\text{[fb]}$ & $f^\text{QCDEW}_\text{NLO}\,\text{[\%]}$ & $\bar{\delta}_\text{EW}\,\text{[\%]}$\\
\hline
{\fontsize{6.0}{6.0}$\text{Unpolarized}$} & $12.745^{+4.9\%}_{-6.2\%}$ & $100$ 
& $12.224^{+5.1\%}_{-6.3\%}$ & $100$ 
& $23.705(1)^{+5.5\%}_{-4.4\%}$ & $100$ 
& $23.184(1)^{+5.6\%}_{-4.5\%}$ & $100$ & $-2.2$\\
\hline
{\fontsize{6.0}{6.0}$W^-_{L}Z_{L}$} & $1.094^{+5.2\%}_{-6.5\%}$ & $8.6$ 
& $1.048^{+5.3\%}_{-6.6\%}$ & $8.6$ 
& $1.407^{+2.6\%}_{-2.1\%}$ & $5.9$ 
& $1.361^{+2.7\%}_{-2.2\%}$ & $5.9$ & $-3.3$\\
{\fontsize{6.0}{6.0}$W^-_{L}Z_{T}$} & $1.508^{+5.8\%}_{-7.0\%}$ & $11.8$ 
& $1.456^{+5.9\%}_{-7.1\%}$ & $11.9$ 
& $3.921^{+7.3\%}_{-5.9\%}$ & $16.5$ 
& $3.869^{+7.4\%}_{-6.0\%}$ & $16.7$ & $-1.3$\\
{\fontsize{6.0}{6.0}$W^-_{T}Z_{L}$} & $1.356^{+5.8\%}_{-7.0\%}$ & $10.6$ 
& $1.347^{+5.8\%}_{-7.0\%}$ & $11.0$ 
& $3.606^{+7.4\%}_{-6.0\%}$ & $15.2$ 
& $3.597^{+7.4\%}_{-6.0\%}$ & $15.5$ & $-0.2$\\
{\fontsize{6.0}{6.0}$W^-_{T}Z_{T}$} & $8.833^{+4.6\%}_{-5.8\%}$ & $69.3$ 
& $8.416^{+4.8\%}_{-5.9\%}$ & $68.8$ 
& $14.664(1)^{+4.7\%}_{-3.8\%}$ & $61.9$ 
& $14.247(1)^{+4.9\%}_{-3.9\%}$ & $61.5$ & $-2.8$\\
\hline
{\fontsize{6.0}{6.0}$\text{Interference}$} & $-0.046(1)^{+6.7\%}_{-5.5\%}$ & $-0.4$ 
& $-0.043(1)^{+6.4\%}_{-5.9\%}$ & $-0.4$ 
& $+0.107(2)^{+16.2\%}_{-16.3\%}$ & $+0.5$ 
& $+0.110(2)^{+15.6\%}_{-15.9\%}$ & $+0.5$ & $+2.8$\\
\hline
\end{tabular}
\caption{\small Unpolarized and doubly polarized cross sections in fb
  together with polarization fractions calculated at LO, NLO EW, NLO
  QCD, and NLO QCD+EW, all in the DPA, in the $WZ$ center-of-mass system for the
  process $p p \to W^- Z\to e^- \nu_e \mu^+\mu^- + X$.   
  The statistical uncertainties (in parenthesis) are given on the last
  digits of the central prediction when significant. Seven-point scale
  uncertainty is also provided for the cross sections as sub- and
  superscripts in percent. In the last column the EW correction relative to the NLO QCD prediction 
  is provided.}
\label{tab:xs_fr}
\end{bigcenter}
\end{table}

In \tab{tab:xs_fr} we present the unpolarized and doubly polarized cross sections (LL, LT, TL, TT) calculated 
using the ATLAS fiducial phase-space volume for the process $p p \to W^- Z\to e^- \nu_e \mu^+\mu^- + X$. 
For the doubly polarized cross sections, polarizations of 
the gauge bosons are defined in the $WZ$ center-of-mass system. 
The interference showing in the bottom row is the difference between the unpolarized cross section and the sum 
of the doubly polarized ones.
To quantify the effect of EW corrections, the relative correction to the LO result is usually used. 
However, since the NLO QCD corrections are large and need to be included in any realistic analyses, 
we therefore define the total EW correction relative to the NLO QCD prediction as
\begin{equation} 
  \bar{\delta}_\text{EW} = (\sigma_\text{NLO}^\text{QCDEW} -
  \sigma_\text{NLO}^\text{QCD})/\sigma_\text{NLO}^\text{QCD},
  \label{eq:deltaEW}
\end{equation} 
to evaluate the effect of NLO EW corrections which are now missing in automated tools. 
This information is shown in the last column.
Polarization fractions, $f$, are calculated as ratios of the 
polarized cross sections over the unpolarized cross section. 
Statistical errors are very small and shown in a few places where they are significant. 
Scale uncertainties are much bigger and are provided for the cross sections as sub- and 
superscripts in percent. These uncertainties are calculated using the seven-point method 
where the two scales $\mu_F$ and $\mu_R$ are varied as 
$n\mu_0/2$ with $n=1,2,4$ and $\mu_0
= (M_W + M_Z)/2$ being the central scale. Additional
constraint $1/2 \leq \mu_R/\mu_F \leq 2$ is used to limit the number of scale
choices to seven at NLO QCD. The cases $\mu_R/\mu_F = 1/4$ or $4$ are excluded, being considered too extreme. 
Note that there are only three possibilities for choosing
$\mu_F$ at LO or NLO EW because of the absence of $\mu_R$. 

At LO, the $W_T Z_T$ is dominant, contributing about $70\%$ to the unpolarized cross section. 
The $W_L Z_T$ and $W_T Z_L$ cross sections are of similar size, about $11\%$ each. The doubly longitudinal 
polarization $W_L Z_L$ cross section amounts to $9\%$, 
which is significant enough for us to hope that it can be measured at ATLAS and CMS. 
The interference is non-vanishing, but very small, being $-0.4\%$. 

At NLO EW level the
$W_L Z_L$, $W_L Z_T$, $W_T Z_L$ and $W_T Z_T$ cross sections are
reduced by $4.2\%$, $3.4\%$, $0.7\%$, and $4.7\%$, 
respectively, with respect to the corresponding LO value.
The unpolarized cross section is reduced by $4.1\%$. 
These numbers are significantly smaller compared to the results found in \cite{Denner:2021csi} for the $ZZ$ process, 
where EW corrections are about $-10\%$ for fully polarized or unpolarized cases. 
This is expected because the photon-quark induced corrections, which are positive, are much larger in the 
$WZ$ process than in the $ZZ$ process as explicitly shown in \cite{Baglio:2013toa}. 
This effect leads to a stronger cancellation between the photon-quark induced corrections and the negative 
virtual corrections in the $WZ$ case, hence leading to smaller total EW corrections.   
 
For the polarization fractions, the
$W_L Z_L$, $W_L Z_T$, and $W_T Z_T$ remain nearly the same as at the LO. 
The $W_T Z_L$ increases slightly by $4\%$. 
The above results show that the EW corrections are very small for integrated quantities.
However, we will see later that EW corrections can be significant for 
transverse momentum distributions in high-energy regions.
 
Unlike EW corrections, NLO QCD corrections to the polarized cross sections are large but not equally
distributed, leading to sizable changes in the 
fractions. In particular, the
$W_L Z_L$, $W_L Z_T$, $W_T Z_L$, and $W_T Z_T$ cross sections increase
by $29\%$, $160\%$, $166\%$, and $66\%$, respectively.
The LL and TT fractions are both reduced to $6\%$ and $62\%$, respectively, 
while the $W_L Z_T$ and $W_T Z_L$ increase to $16\%$ each. 
It is unfortunate that QCD corrections reduce the LL fraction, but, luckily the value is still large enough to be measured. 
Before comparing our results to the new ATLAS measurement, we notice that 
the polarization fractions at NLO QCD+EW level are slightly different between the $W^- Z$ and $W^+ Z$ processes. 
We recall the $W^+ Z$ fractions \cite{Le:2022lrp}, $5.6\%$ (LL), $15.6\%$ (LT), $15.1\%$ (TL), and $63.0\%$ (TT), which 
are to be compared with the ones in \tab{tab:xs_fr} for the $W^- Z$ case.   

A full comparison between our NLO QCD+EW predictions to the new
preliminary ATLAS results \cite{ATLAS:2022conf53}  for both the $W^+
Z$ and $W^- Z$ processes is shown in \tab{tab:compare_ATLAS}. The
errors on our NLO QCD+EW predictions are calculated only from the
scale uncertainties, taking the averaged value of the two errors (at
the cross sections) to get a symmetric result. The discrepancy between
our NLO QCD+EW polarization fractions and  the ATLAS results is
quantified by the pull defined as
$(f_{\text{th}}-f_{\text{exp}})/\sigma$ where
$\sigma=\sqrt{\sigma_{th}^2 +\sigma_{exp}^2 }$. The doubly transverse
polarized fractions, being largest and most precisely measured, are in
good agreement with the discrepancy being less than $0.5$ standard
deviations for both processes. The doubly longitudinal fractions,
being smallest, are also in good agreement, within $1$ standard
deviation. However, the experimental uncertainties are large, being $22\%$
($25\%$) for the $W^+ Z$ ($W^- Z$) processes. The largest deviations
are found for the $W_L^- Z_T$ and $W_T^- Z_L$ fractions where the
magnitudes of the pulls are of $1.3$ for both cases. Though this good
agreement is encouraging, one must pay attention to the experimental
uncertainties, which range from $6\%$ to $36\%$, while the theory
uncertainties are from $5\%$ to $8\%$. To reach the level of precisepolarization measurements in di-boson productions, further work is
needed from both the theory and experimental sides.
\begin{table}[th!]
 \renewcommand{\arraystretch}{1.3}
\begin{bigcenter}
\begin{tabular}{|c|c|c|c|}\hline
  & NLO QCD+EW & ATLAS & Pull\\
\hline
{$W^+_{L}Z_{L}$} & $0.056 \pm 0.003$ & $0.072 \pm 0.016$ & $-1.0$\\
{$W^+_{L}Z_{T}$} & $0.156 \pm 0.013$ & $0.119 \pm 0.034$ & $+1.0$\\
{$W^+_{T}Z_{L}$} & $0.151 \pm 0.012$ & $0.153 \pm 0.033$ & $-0.1$\\
{$W^+_{T}Z_{T}$} & $0.630 \pm 0.041$ & $0.660 \pm 0.040$ & $-0.5$\\
\hline
{$W^-_{L}Z_{L}$} & $0.059 \pm 0.003$ & $0.063 \pm 0.016$ & $-0.3$\\
{$W^-_{L}Z_{T}$} & $0.167 \pm 0.014$ & $0.11 \pm 0.04$ & $+1.3$\\
{$W^-_{T}Z_{L}$} & $0.155 \pm 0.013$ & $0.21 \pm 0.04$ & $-1.3$\\
{$W^-_{T}Z_{T}$} & $0.615 \pm 0.041$ & $0.62 \pm 0.05$ & $-0.1$\\
\hline
\end{tabular}
\caption{\small Comparison with ATLAS measurements
  \cite{ATLAS:2022conf53}. The pull is defined as
  $(\text{Theory}-\text{Experiment})/\sigma$ where $\sigma$ is the
  combined error calculated in quadrature.}
\label{tab:compare_ATLAS}
\end{bigcenter}
\end{table}
\subsection{Kinematic distributions}
\label{sect:dist}
We now discuss kinematic distributions. 
In order to facilitate comparison between the two processes $W^+ Z$
and $W^- Z$, we first present in this section the same set of plots as
in \cite{Le:2022lrp} but for the $W^- Z$. As expected, the results for
the $W^- Z$ case look very similar to the $W^+ Z$ case except for the
absolute values of the cross sections. By comparing the figures of the
two papers, the reader will find the shapes of the distributions very
much alike. At the end of the section, a new interesting distribution
of the rapidity separation between the electron and the $Z$ boson will
be presented for both processes.

\begin{figure}[t!]
  \centering
  \begin{tabular}{cc}
  \includegraphics[width=0.48\textwidth]{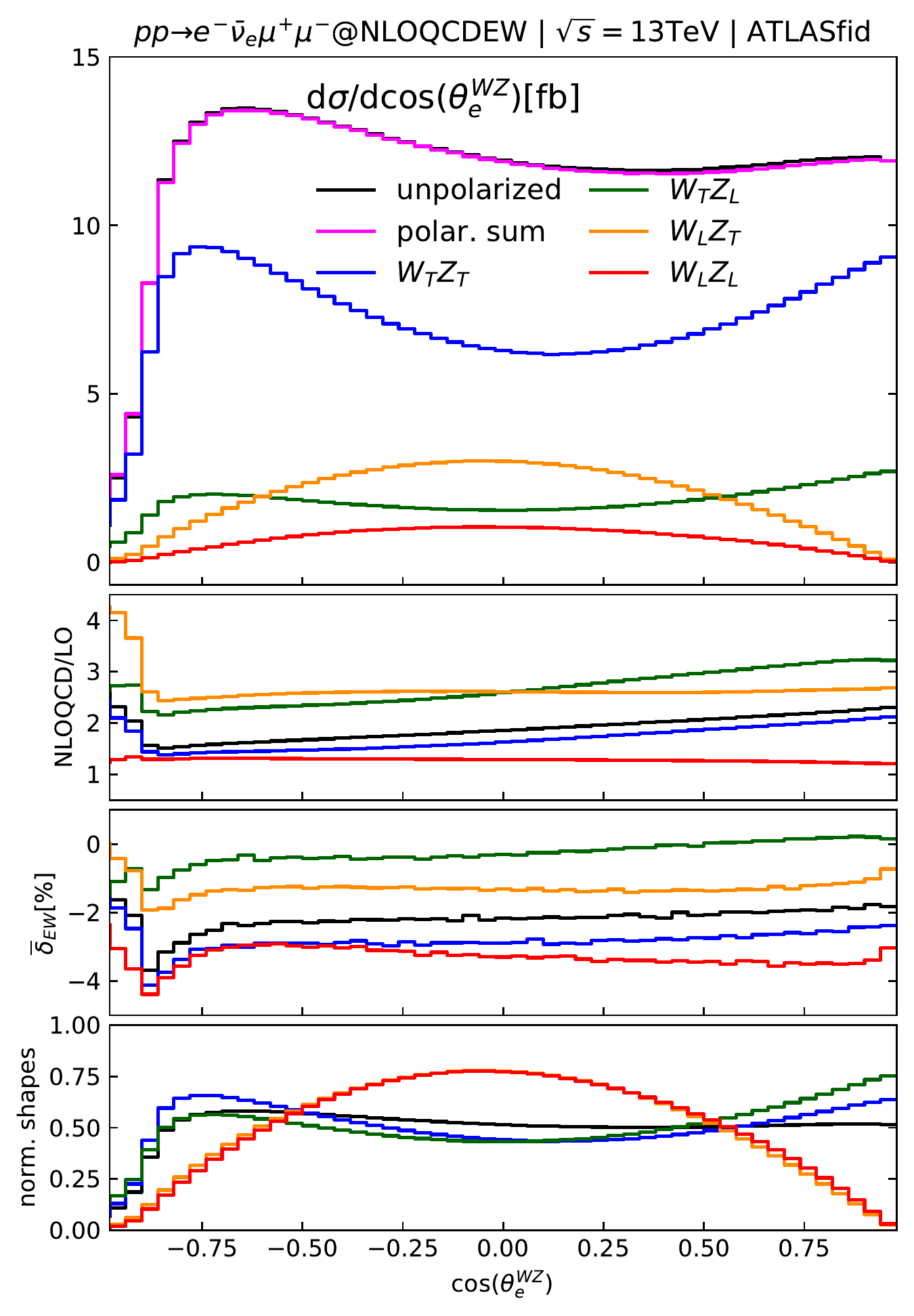} 
  \includegraphics[width=0.48\textwidth]{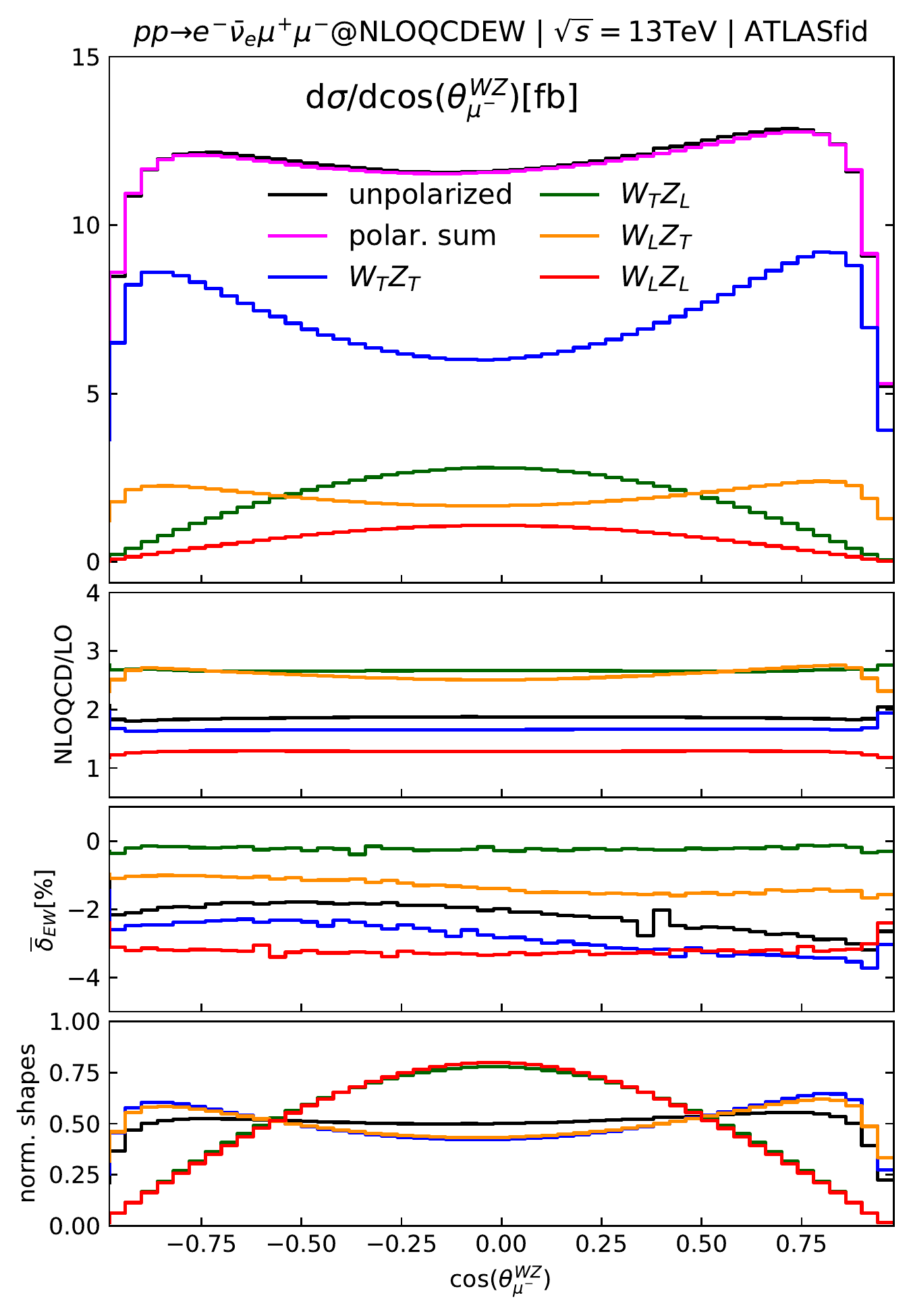}
  \end{tabular}
  \caption{Distributions in $\cos\theta^{WZ}_{e}$ (left) and
    $\cos\theta^{WZ}_{\mu^-}$ (right). These angles are calculated in
    the $WZ$ center-of-mass system (more details are provided in the
    text), hence denoted with the $WZ$ superscript. The big panel
    shows the absolute values of the cross sections at NLO QCD+EW. The
    middle-up panel displays the ratio of the NLO QCD cross sections
    to the corresponding LO ones. The middle-down panel shows
    $\bar{\delta}_{\text{EW}}$, the EW corrections relative to the NLO
    QCD cross sections, in percent. In the bottom panel, the
    normalized shapes of the distributions are plotted to highlight
    differences in shape.}
  \label{fig:dist_costheta}
\end{figure}
In \fig{fig:dist_costheta} we present the differential cross sections
in $\cos\theta^{WZ}_{e}$ (left) and $\cos\theta^{WZ}_{\mu^-}$
(right). The polar angle $\theta^{WZ}_{\ell}$ is defined as the angle
between the momentum of the parent gauge boson calculated in the $WZ$
c.m.s ($\vec{p}^\text{WZ-cms}_V$) and the momentum of the lepton
calculated in the gauge boson rest frame
($\vec{p}^\text{V-rest}_{\ell}$). These angles are chosen for our
analyses because the electron angle distribution is sensitive to the
$W$ boson polarizations, while the muon angle distribution is
sensitive to the $Z$ boson polarizations. 

In the top panels, we display the NLO QCD+EW differential cross
sections for the  double polarizations LL (red), LT (orange), TL
(green), TT (blue). Their sum is plotted in magenta (only shown in the
top panels), while the unpolarized cross section is in black. The
difference between the  unpolarized and the polarization sum is the
interference shown in the last row of \tab{tab:xs_fr}. 
As seen from these plots, the interference effect is negligible across
the full range of the angle for both cases. Similar to the case of
$W^+ Z$, the $W_T^- Z_T$ cross section is largest while the $W_L^-
Z_L$ smallest for both distributions. Comparing the $W_L^- Z_T$ to the
$W_T^- Z_L$ $\cos\theta^{WZ}_{e}$ distribution, we see that the
transverse $W$ mode is more dominant at the edge regions where
$|\cos\theta^{WZ}_{e}| \approx 1$, while the longitudinal $W$ cross
section is larger in the center region $|\cos\theta^{WZ}_{e}| <
0.5$. The same features are observed in the right plot of the muon
angle. The depletion at $|\cos\theta^{WZ}_{\mu^-}| \approx 1$ in the
$W_T Z_T$ and $W_L Z_T$ distributions in the right plot is due to the
$p_T$ and $\eta$ cuts on the muon and anti-muon. These cuts do not
affect the shapes of the $W_T Z_L$ and $W_L Z_L$ polarizations as
observed in \cite{Denner:2020eck}.  Similar results are seen in the
left plot, but there is no depletion at $\cos\theta^{WZ}_{e} \approx
+1$ because there are no $p_T$ and $\eta$ cuts on the neutrino.  
\begin{figure}[t!]
  \centering
  \begin{tabular}{cc}
  \includegraphics[width=0.48\textwidth]{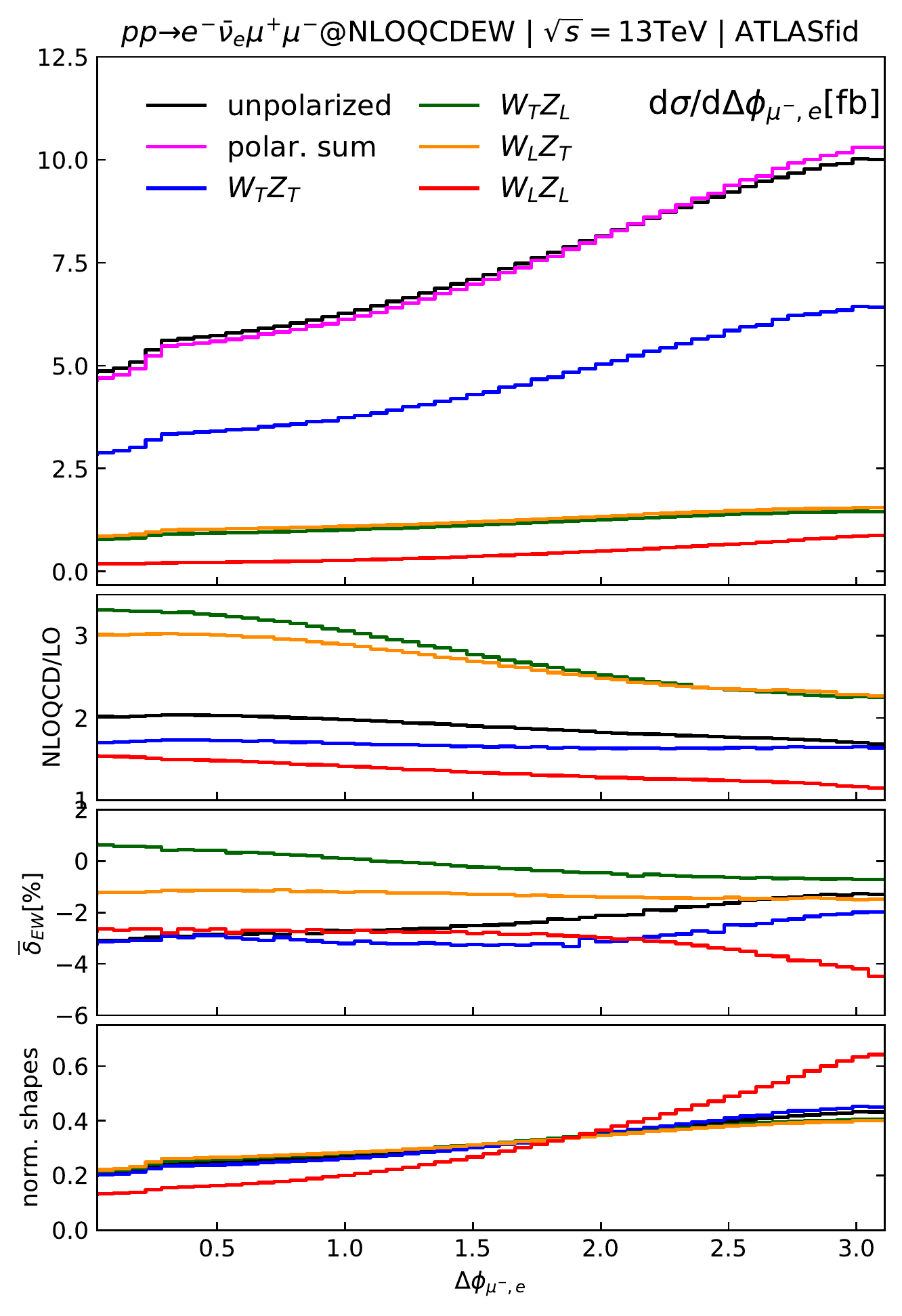}& 
  \includegraphics[width=0.48\textwidth]{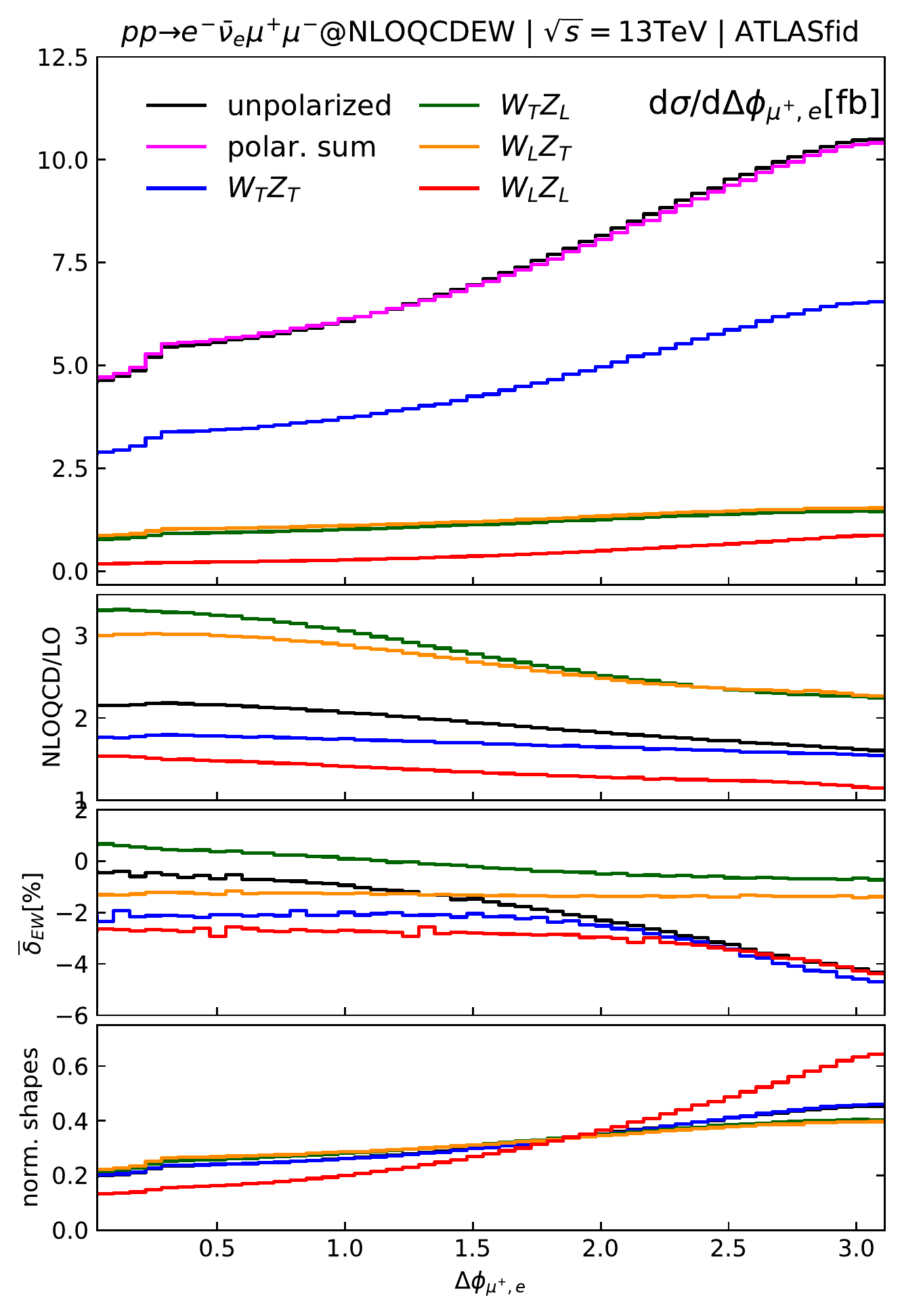}
  \end{tabular}
  \caption{Same as \fig{fig:dist_costheta} but for the azimuthal
    separation between the electron and the muon
    $\Delta\phi_{\mu^-,e^-}$ (left) and between the electron and the
    anti-muon $\Delta\phi_{\mu^+,e^-}$ (right).}
  \label{fig:dist_phi}
\end{figure}

The ratios of the NLO QCD cross section to the corresponding LO one
($K$-factor) are plotted in the middle-up panel, while the
$\bar{\delta}_\text{EW}$ corrections are shown in the middle-down
panel. The QCD $K$-factor for the $LL$ polarization is the smallest
(about $1.3$) and rather flat in the whole range of
$\cos\theta^{WZ}_{e}$ while it is larger for the other polarizations,
with a sharp rise in the region $\cos\theta^{WZ}_{e}<-0.8$ where there
is a stronger depletion of the differential cross section. The
$K$-factor is greater than $4$ at $\cos\theta^{WZ}_{e}=-1$ for the
$W_L Z_T$ case. For the muon angle distributions, the QCD $K$-factors
are flatter in the whole range, varying from $1.2$ to $2.8$. In the EW
correction panels, note that $\bar{\delta}_\text{EW}$ is defined with
respect to the NLO QCD result, see \eq{eq:deltaEW}. The
$|\bar{\delta}_\text{EW}|$ is smallest for the TT component and
largest for the LL component. They remain small in the whole range of
$\cos\theta^{WZ}_\ell$ and vary from $-4.2\%$ to $0.1\%$.

Finally, in the bottom panels we show the distributions
$d\sigma/d\cos\theta$ normalized to their corresponding integrated
cross sections presented in \tab{tab:xs_fr}. This helps us to see the
differences in shape of the various polarizations. From the
$\cos\theta^{WZ}_{e}$ distribution, we observe two distinct shapes:  
The $W_L Z_L$ and $W_L Z_T$ have the same shape with a maximum at
$\cos\theta^{WZ}_{e} = 0$, while the $W_T Z_L$ and $W_T Z_T$ are
similar with a minimum at the center. This feature is well expected
because the $\cos\theta^{WZ}_{e}$ distribution is sensitive to the
polarizations of the $W$ boson and is not much affected by the
polarization of the $Z$ boson. The same thing can be said for the
$\cos\theta^{WZ}_{\mu^-}$ case where the longitudinal and transverse
$Z$ bosons produce two different shapes.
\begin{figure}[t!]
  \centering
  \begin{tabular}{cc}
  \includegraphics[width=0.48\textwidth]{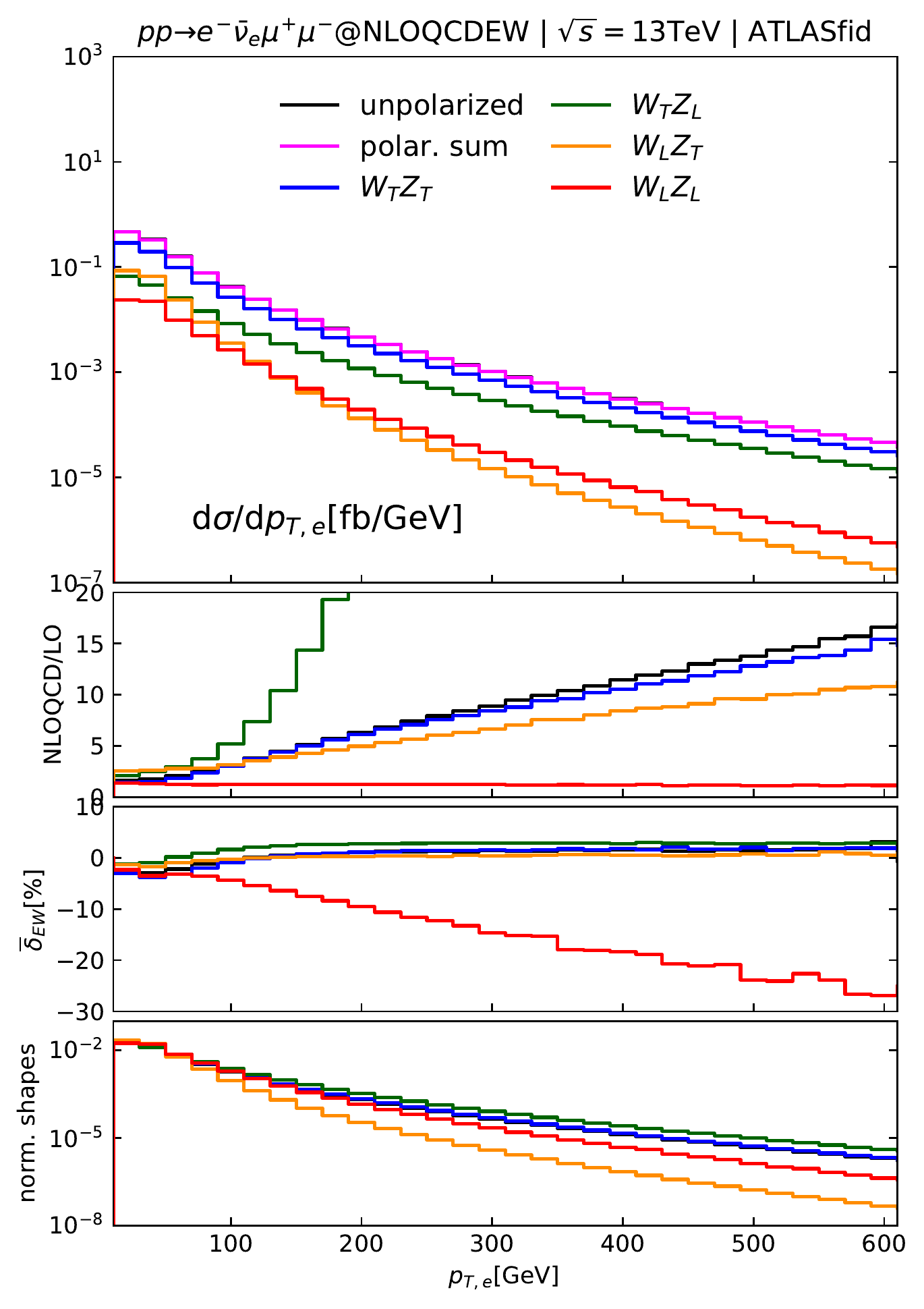}& 
  \includegraphics[width=0.48\textwidth]{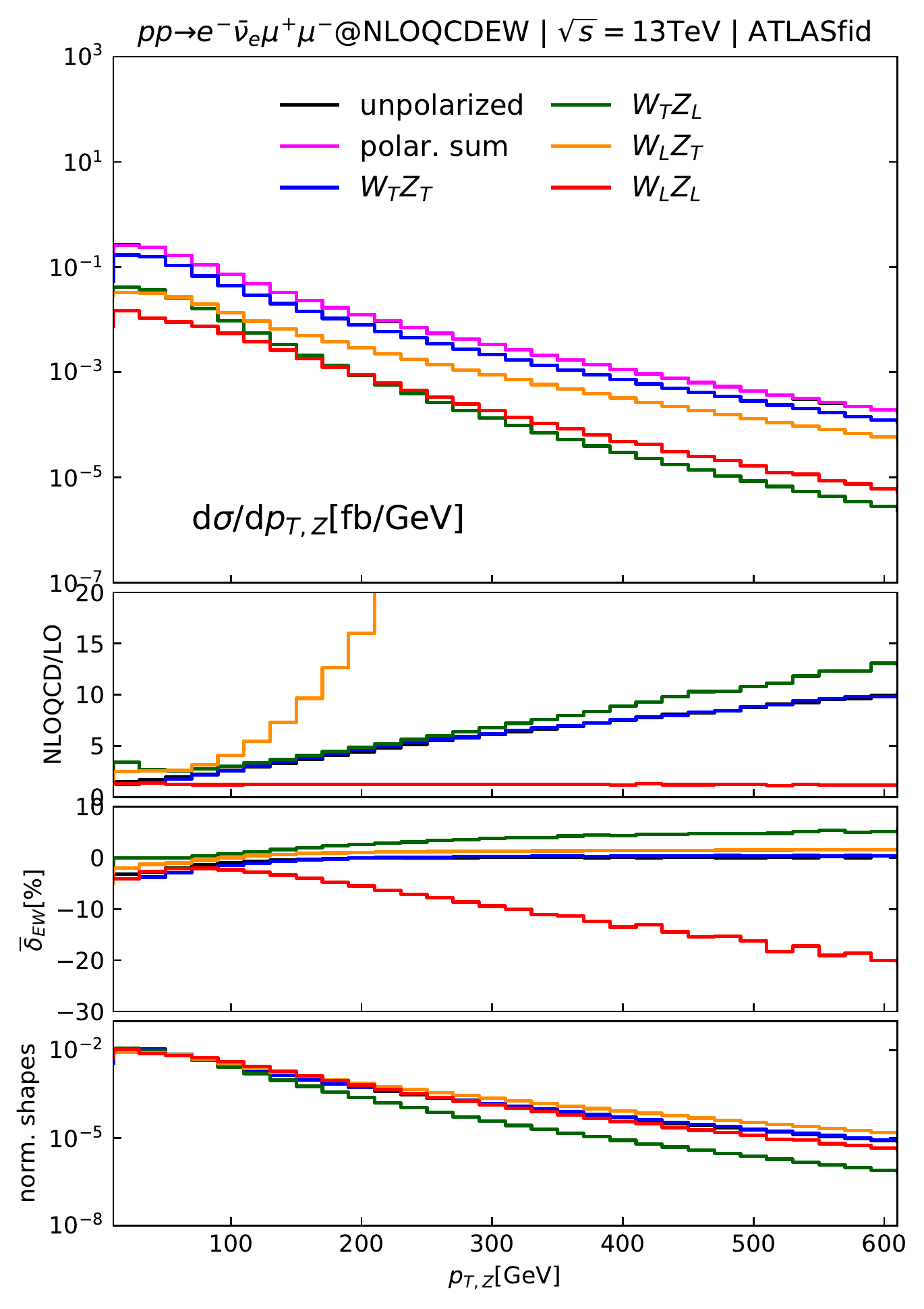}
  \end{tabular}
  \caption{Same as \fig{fig:dist_costheta} but for the transverse momentum of the 
electron (left) and the $Z$ boson (right).}
  \label{fig:dist_pT}
\end{figure}

In the same format and color code, we plot in \fig{fig:dist_phi} the
distributions in the azimuthal separation between the electron and the
muon $\Delta\phi_{\mu^-,e^-}$ (left), between the electron and the
anti-muon $\Delta\phi_{\mu^+,e^-}$ (right). All polarizations show an
increase in the differential cross section with increasing separation,
however the QCD $K$-factors decrease.
As in the case of the $\cos\theta^{WZ}_{\ell}$ distributions, the QCD
corrections are large while the EW corrections are small, being less
than $5\%$ across the full range of $\Delta\phi \in [0,\pi]$. The
normalized shape panels show that this distribution can give extra
power to separate the LL polarization as it has a different shape from
the other polarizations for both the left and the right plots.

The transverse momentum distributions of the electron (left) and of
the $Z$ boson (right) are presented in \fig{fig:dist_pT}. More clearly
than the above angular distributions, these distributions show that
the QCD and EW corrections are not the same for different
polarizations. QCD correction is largest for the $W_T Z_L$ in the
$p_{T,e}$ distribution, while the $W_L Z_T$ is largest in the
$p_{T,Z}$ one. EW correction is largest for the LL polarization for
both distributions. The correction is negative and its magnitude
increasing with energy due to the double- and single-Sudakov
logarithms in the virtual contribution. Compared to the NLO QCD
prediction, the EW correction is $-10\%$ at around $200$~GeV and
reaching $-27\%$ at $600$~GeV for the $p_{T,e}$ distribution. For the
$p_{T,Z}$ distribution, the correction is smaller, being $-5\%$ at
$200$~GeV and $-20\%$ at $600$~GeV. The normalized shapes of the four
polarizations are indistinguishable for $p_T < 100$~GeV and become
more diverged at large values.
\begin{figure}[t!]
  \centering
  \begin{tabular}{cc}
  \includegraphics[width=0.48\textwidth]{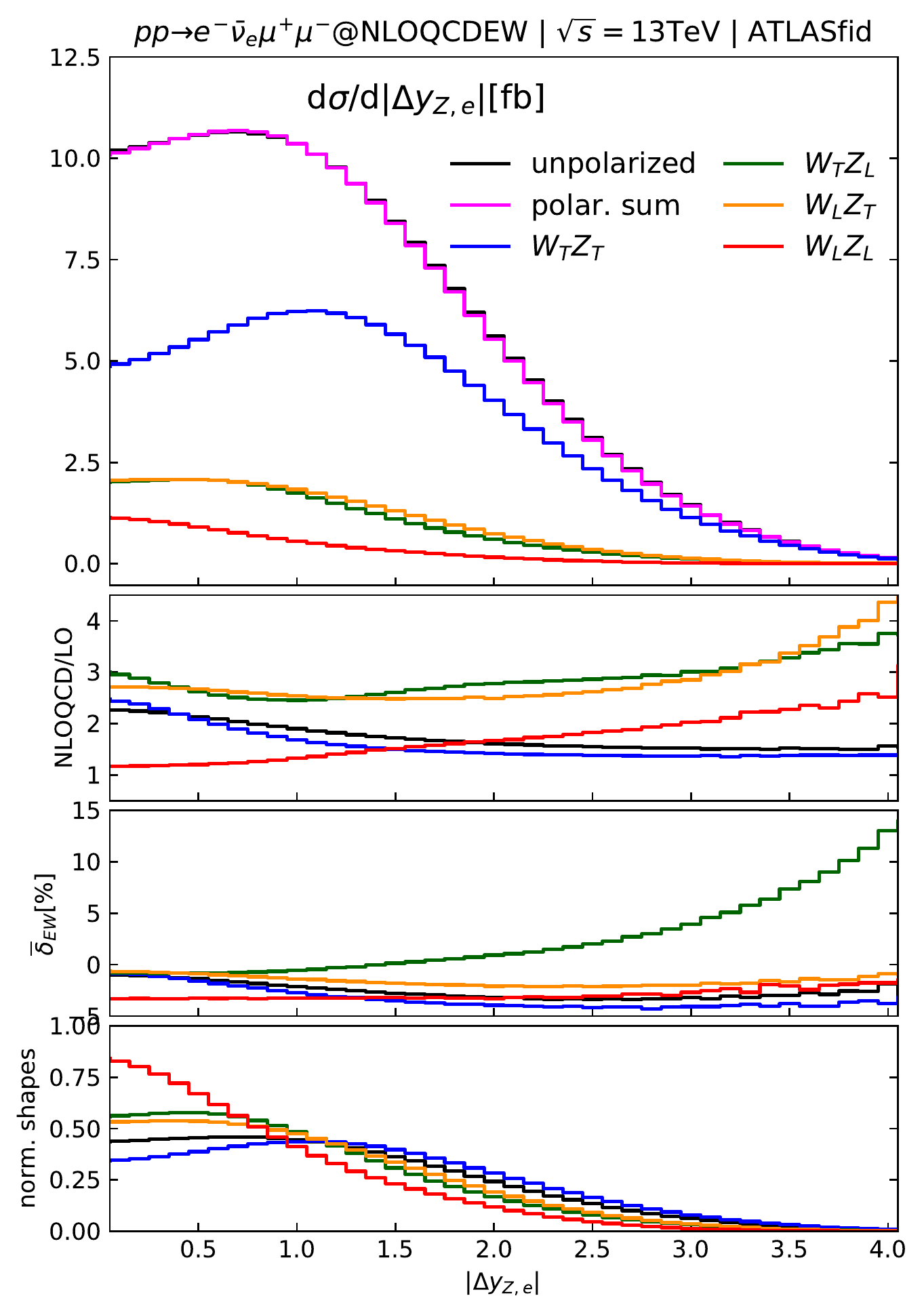}& 
  \includegraphics[width=0.48\textwidth]{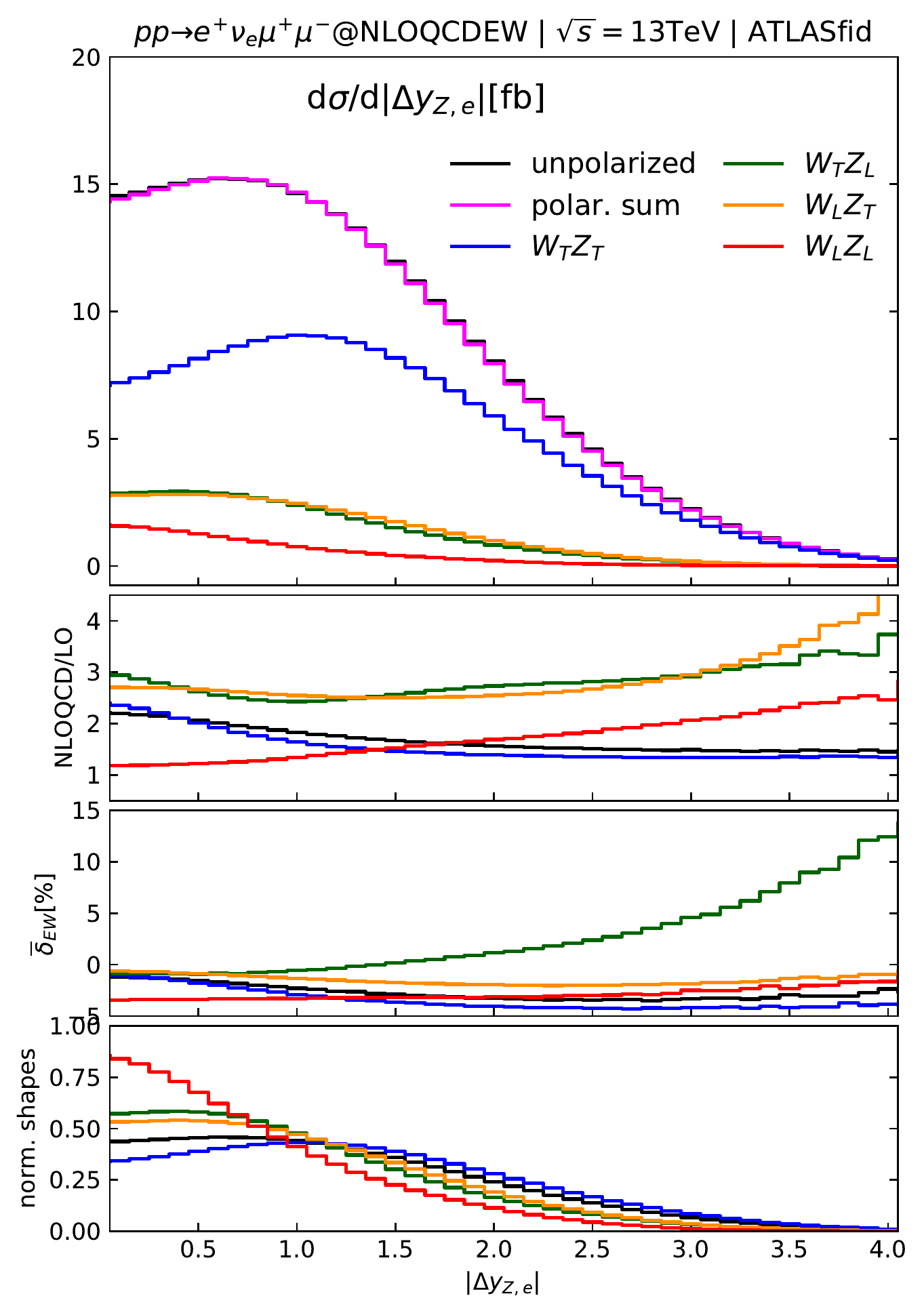}
  \end{tabular}
  \caption{Same as \fig{fig:dist_costheta} but for the rapidity
    separation (in absolute value) between the electron/positron and
    the $Z$ boson. The left plot is for the process $W^- Z$ while the
    right plot is $W^+ Z$.}
  \label{fig:dist_Dy_eZ}
\end{figure}

Another interesting distribution which can help to distinguish the
doubly longitudinal polarization is the rapidity separation between
the electron and the $Z$ boson. This is plotted in
\fig{fig:dist_Dy_eZ} for both the $W^- Z$ (left) and $W^+ Z$ (right)
processes. The latter was not shown in \cite{Le:2022lrp}, hence it is
here presented for the sake of comparison. The normalized shape panels
show clearly that the LL polarization is different from the other
cases. Another remarkable feature is the large EW correction, which
increases with large rapidity separation, in the $W_T Z_L$
polarization. Since the photon-quark induced processes are separated
from the photon-radiation processes in our calculation, we can
investigate the origin of this large EW correction. We found that it
is due to the photon-quark induced processes with an extra jet in the
final state. The jet allows for new kinematic configurations such as
the jet recoiling against a hard $W$ boson leaving the $Z$ boson the
freedom to be soft, or the hard jet recoiling against the $Z$ boson
while the $W$ is soft. As shown in \cite{Baglio:2013toa}, these kinds
of configurations can lead to large corrections proportional to
$\alpha \log^2(p_{T,jet}^2/M_V^2)$ ($V=W,Z$). This argument also holds
for the gluon-quark induced processes in the QCD
corrections. \fig{fig:dist_Dy_eZ} indeed shows that the QCD
corrections are large at large rapidity separation for the $W_T Z_L$
and $W_L Z_T$ polarizations. Turning off the gluon-quark induced
processes makes this correction significantly smaller. It is
interesting to note that while the gluon-quark induced processes
affect both the $W_T Z_L$ and $W_L Z_T$ polarizations, the
photon-quark induced processes increase only the $W_T Z_L$ case. This
reminds us of the difference between QCD and EW corrections.

\section{Conclusions}
\label{sect:conclusion}
In this paper we have studied the doubly-polarized production of
$W^\pm Z$ pairs in fully-leptonic channels at the LHC at NLO accuracy
for both the QCD and EW corrections. Numerical results for the $W^+ Z$
process were already presented in the short letter \cite{Le:2022lrp}. 
Here we present a detailed description of the calculation 
method behind the results of \bib{Le:2022lrp} and provide further
numerical results, mainly for the $W^- Z$ process.
  
The method to calculate NLO QCD corrections for doubly-polarized cross
sections in di-boson productions has been established in
\cite{Denner:2020bcz} using the double-pole approximation. NLO EW
corrections are more complicated and have been recently calculated in
\cite{Denner:2021csi} for the $ZZ$ process, also in the DPA. In this
work, we have extended this method to cover the case of a charged
current, namely the $WZ$ process. The method described here can also
be straightforwardly used for the $W^+W^-$ process.

In the numerical result section, we have presented new results for the $W^-
Z$ process at the NLO QCD+EW level. This has bee  done in such a way
that the reader can easily compared to the corresponding $W^+ Z$
results provided in \bib{Le:2022lrp}. We note that this is also the
first time NLO QCD results for the $W^- Z$ process are presented as
\bib{Denner:2020eck} published only the $W^+ Z$ NLO QCD results. 

Integrated doubly-polarized cross sections have been calculated
together with seven-point scale uncertainties. A comparison between
our predictions and the new ATLAS measurement~\cite{ATLAS:2022conf53}
has been tabulated and discussed, showing very good agreement within
$1.5$ standard deviations. However, the experimental precision is
still limited, at the level of tens of percents. 

We have presented also differential distributions mainly for the $W^-
Z$ process together with a detailed analysis of the results. New
distributions of the rapidity separation between the electron and the
$Z$ boson ($\Delta y_{Z,e}$) have been shown for both $W^- Z$ and $W^+
Z$ processes. The kinematic variables chosen here are those which
provide discrimination power to separate different polarizations
(mostly lepton angular observables) or those which show the importance
of EW corrections (lepton $p_T$). We have found that, as in the case
of the $W^+ Z$ channel, EW corrections are most sizable in the
$p_{T,e}$ and $p_{T,Z}$ distributions of the doubly-longitudinal
polarization in the $W^- Z$ channel. It is also found that the
rapidity separation $\Delta y_{Z,e}$ can help to single out the LL
polarization.
\acknowledgments

This research is funded by the Vietnam National Foundation for Science
and Technology Development (NAFOSTED) under grant number
103.01-2020.17.


\providecommand{\href}[2]{#2}\begingroup\raggedright\endgroup
\end{document}